\documentclass[preprint,preprintnumbers,nofootinbib,aps,10pt,twocolumn]{revtex4-1}
\usepackage{amsmath,amssymb,bm,epsfig}
\usepackage{color}
\usepackage{natbib}
\usepackage{hyperref} 
\usepackage{ulem}
\usepackage{graphicx}
\usepackage{xcolor,colortbl}
\RequirePackage{lineno}
\newcommand{\nn}{\nonumber}
\newcommand{\sNN}{\sqrt{s_{\textrm{NN}}}}

\definecolor{Gray}{gray}{0.85}
\newcolumntype{a}{>{\columncolor{Gray}}c}

\def \beq{\begin{equation}}
\def \eeq{\end{equation}}
\def \beqa{\begin{eqnarray}}
\def \eeqa{\end{eqnarray}}
\def \la{\langle}
\def \ra{\rangle}
\def \l{\left(}
\def \r{\right)}
\def \l{\left(}
\def \r{\right)}
\def \d{$D^0$ }
\def \dbar{$\overline{D^0}$ }
\def \dbare{$\overline{D^0}$}
\def \vavg{$v_1^{\text{avg}}$ }
\def \vdiff{$v_1^{\text{diff}}$ }

\def \vdiffe{$v_1^{\text{diff}}$}
\def \svdiff{$dv_1^{\text{diff}}/d\eta$ }
\def \svavg{$dv_1^{\text{avg}}/d\eta$ }
\def \svdiffe{$dv_1^{\text{diff}}/d\eta$}

\begin{document}

\title{Interplay of drag by hot matter  and electromagnetic force on the 
directed flow of heavy quarks}

\author{Sandeep Chatterjee$^{1,2}$}
\email{Sandeep.Chatterjee@fis.agh.edu.pl}
\author{Piotr Bo{\.z}ek$^{1}$}
\email{Piotr.Bozek@fis.agh.edu.pl}
\affiliation{$^1$AGH University of Science and Technology,\\ 
Faculty of Physics and Applied Computer Science,\\
aleja Mickiewicza 30, 30-059 Krakow, Poland}
\affiliation{$^2$Department of Physical Sciences,\\
Indian Institute of Science Education and Research Berhampur,\\ 
Transit Campus (Govt ITI), Berhampur-760010, Odisha, India}

\begin{abstract}

Rapidity-odd directed flow in heavy ion collisions can originate from two very 
distinct sources in the collision dynamics i. an initial tilt of the fireball in the reaction 
plane that generates directed flow of the constituents independent of their charges, 
and ii. the Lorentz force due to the strong primordial electromagnetic field that 
drives the flow in opposite directions for constituents carrying unlike sign charges. 
We study the directed flow of open charm mesons $D^0$ and $\overline{D^0}$ in the 
presence of both these sources of directed flow. The drag from the tilted matter  dominates over the Lorentz force resulting 
in same sign flow for both $D^0$ and $\overline{D^0}$, albeit of different magnitudes.
Their average directed flow is about ten times larger than their difference. This charge
splitting in the directed flow is a sensitive probe of the electrical conductivity 
of the produced medium. We further study their beam energy dependence; while the average directed flow shows a decreasing 
trend, the charge splitting 
remains flat from $\sqrt{s_{NN}}=60$~GeV to $5$~TeV.


\end{abstract}

\maketitle

A strongly interacting medium is expected to be formed in relativistic heavy ion collisions. Transport coefficients of the dense matter 
are one of the foremost indicators of the nature of the relevant 
degrees of freedom that constitute this medium. Shear and bulk viscosities which 
are the transport coefficients corresponding to the energy momentum tensor has been extensively 
studied and extracted from data leading to considerable understanding of the nature of the strongly 
interacting quark gluon plasma that is expected to be created in these collisions~\cite{Ollitrault:2010tn, 
*Heinz:2013th,*Gale:2013da}. The electric conductivity $\sigma$ is the transport coefficient corresponding 
to the electric charge. An estimate of $\sigma$ in heavy-ion collisions will further add to our understanding of 
the medium properties of hot and dense QCD matter~\cite{Gupta:2003zh,*Hirono:2012rt,*Cassing:2013iz,*Yin:2013kya,
*Finazzo:2013efa,*Puglisi:2014sha,*Greif:2014oia,*Srivastava:2015via,*Ghosh:2016yvt,*Hattori:2016cnt,*Feng:2017tsh,
*Thakur:2017hfc,*Mitra:2017sjo,*Ghosh:2018kst,Ding:2010ga,*Amato:2013naa,*Brandt:2012jc}. 
Further, in the light of attempts to calibrate the magnitude and temporal dependence of the electromagnetic 
(EM) field produced in heavy-ion collisions~\cite{Kharzeev:2007jp,*Tuchin:2010vs,*Bzdak:2011yy,*Voronyuk:2011jd,*Deng:2012pc} 
and its phenomenological consequences like the chiral magnetic effect~\cite{Fukushima:2008xe}, the knowledge 
of $\sigma$ is of utmost importance.

Heavy quarks (HQs) by virtue of being several times more massive than the highest ambient temperatures 
achieved in a collision are expected to be produced only in primordial collisions. 
Thus, they serve as excellent 
probes that witness the  spacetime evolution of the fireball~\cite{Rapp:2009my,*Andronic:2015wma,
*Aarts:2016hap}. Charged HQs are formed early and their deflection by the Lorentz force probes the EM fields at the very early stage of the collision. In the following
as heavy quarks we study specifically  charm and anticharm quarks, observed in the final state in open charm mesons \d and \dbar.

The initial state of a non-central heavy-ion collision is expected to break the forward-backward symmetry by a tilt of the fireball away from the beam axis~\cite{Csernai:1999nf,
Snellings:1999bt,Lisa:2000ip,Adil:2005qn,Bozek:2010bi,Steinheimer:2014pfa}. 
This is confirmed by the observation of rapidity-odd directed flow $v_1$ of charged particles~\cite{Back:2005pc,
Abelev:2008jga,Abelev:2013cva,Singha:2016mna}. On the other hand,
 HQs which are produced according to the 
profile of the binary collision sources are distributed symmetrically in the the forward-backward direction. 
At nonzero rapidities it results in a  shift of the HQ production points 
 from the tilted bulk.
 Recently, within the 
framework of Langevin dynamics coupled to a hydrodynamic background, it has been shown that this difference 
 between the bulk matter and the HQ production points can lead to HQ $v_1$ that is of same sign as the 
bulk but several times larger~\cite{Chatterjee:2017ahy}. Similar trends are also expected from a transport 
model approach~\cite{Nasim:2018hyw}. Such large HQ $v_1$ compared to the charged particle $v_1$ is a clear 
signature of the tilt of initial source.

A rapidity-odd $v_1$ can also arise due to the presence of  EM field~\cite{Gursoy:2014aka,Das:2016cwd}. However, unlike 
the $v_1$ sourced by the expansion of the tilted fireball which is of same sign for both \d and \dbare~\cite{Chatterjee:2017ahy}, the Lorentz force experienced 
by charm and anti-charm quarks being in opposite direction, the resulting $v_1$ is of opposite sign for \d and 
\dbare~\cite{Das:2016cwd}. In this work, we calculate the directed flow coefficient $v_1$ of \d and \dbar mesons under the combined influence 
of the drag from the tilted source and the EM fields.

The  forward-backward asymmetry of the  initial fireball can originate from
an asymmetric deposition of entropy from forward and backward going 
participants~\cite{Brodsky:1977de,Bialas:2004su,Adil:2005qn}.
Such an
  ansatz in which a participant is postulated to deposit entropy 
preferably along its direction of motion, has been successful is describing the observed charged particle 
directed flow~\cite{Bozek:2010bi}.

The initial density $s\l\tau_0,x,y,\eta_{||}\r$  in the   Glauber model with asymmetric entropy deposition can be written as~\cite{Bozek:2010bi}
\beqa
 s\l\tau_0,x,y,\eta_{||}\r &=& s_0 \left[ \l1-\alpha\r\l N_{part}^+ f_+\l\eta_{||}\r+\right.\right.\nn\\
 && \left.\left.N_{part}^- f_-\l\eta_{||}\r\r + \alpha N_{coll} \right]f\l\eta_{||}\r\label{eq.ic}
\eeqa
where $N_{part}^+$ and $N_{part}^-$ are the densities  of participant sources from 
the forward and backward going nuclei respectively evaluated at $\l x,y\r$ and $N_{coll}$ is the density  of binary collisions. $\tau=\sqrt{t^2-z^2}$ is the proper time and $\eta_{||}=\frac{1}{2}\log\frac{\l t+z\r}{\l t-z\r}$ 
is the spacetime rapidity. $\tau_0$ is the proper time when the HQ starts to interact with the bulk and also the 
initial proper time to start the hydrodynamic evolution. In principle these time scales could be different and there have 
been previous studies on the preequilibrium dynamics of the HQ~\cite{Chesler:2013urd,*Das:2015aga}. However, in this 
first study of the combined effect of drag and EM field, we work with the simple ansatz that the HQ interaction with 
the medium starts at the same time as the hydrodynamic expansion of the bulk.
$f\l\eta_{||}\r$ is the rapidity-even profile
\beqa
 f\l\eta_{||}\r &=& \exp\l-\theta\l|\eta_{||}|-\eta_{||}^0\r\frac{\l|\eta_{||}|-\eta^0_{||}\r^2}{2\sigma_\eta^2}\r 
 \label{eq.feta}
\eeqa
while the tilt is introduced via the factors $f_{+,-}\l\eta_{||}\r$
\beq
 f_+\l\eta_{||}\r = \left\{\begin{array}{lr}
                            1, & \eta_{||}>\eta_T\\
                            \frac{\eta_T+\eta_{||}}{2\eta_T}, & -\eta_T\leq\eta_{||}\leq\eta_T\\
                            0, & \eta_{||}<-\eta_T
                           \end{array}\right.
\label{eq.fpm}
\eeq
with  $f_-\l\eta_{||}\r=f_+\l-\eta_{||}\r$.
A suitable choice for $s_0$, $\eta^0_{||}$, $\alpha$ and $\sigma_\eta$ are made to reproduce  
the  charged particle distribution in pseudorapidity at different centralities.  Finally, $\eta_T$ is adjusted to reproduce the 
observed  rapidity-odd directed flow of charged particles. 

All our results are for the  $0-80\%$ centrality bin. This corresponds to a choice of impact 
parameter, $b=8.3$ fm within our optical Glauber model approach to obtain the initial condition. The 
$\l 3+1\r$-dimensional relativistic hydrodynamic evolution are carried out by the publicly available vHLLE 
code~\cite{Karpenko:2013wva}. The freezeout hypersurface is assumed to be at a constant temperature $T=150$ MeV, where statistical emission of hadrons  happens 
\cite{Chojnacki:2011hb}. Details of the model and parameters  of the hydrodynamic model 
used at the Brookhaven Relativistic Heavy Ion Collider (RHIC) and at the CERN Large Hadron Collider (LHC) energies can be found in \cite{Bozek:2011ua,*Bozek:2012qs}.

The full spacetime history of the flow velocity and $T$ fields obtained from the hydrodynamic evolution 
are fed as input to the  Langevin dynamics  of the HQs
\beqa
\Delta { r}_i &=& \frac{{ p}_i}{E}\Delta t\label{eq.Langevinx}\\
\Delta { p}_i &=& -\gamma { p}_i\Delta t + \rho_i\sqrt{2D\Delta t} + {F}^{EM}_i\label{eq.Langevinp}
\eeqa
where ${\bf F}^{EM}$ refers to the Lorentz force due to the EM field. The updates of the position and momentum 
vectors of the HQ in time interval $\Delta t$ are denoted by $\Delta {\bf r}$ and $\Delta {\bf p}$ respectively. 
Here $i=x$, $y$ and $z$ are the three Cartesian coordinate components. The initial position coordinates are sampled 
from the binary collision profile while the momenta are generated from p+p events of PYTHIA~\cite{Sjostrand:2006za,*Sjostrand:2007gs}. 
The HQ interaction with the medium is encoded in the drag $\gamma$ and diffusion $D$ coefficients. $\rho_i$ satisfy $\la\rho_i\ra=0$ and 
$\la\rho_i\rho_j\ra=\delta_{ij}$ which is realized by randomly sampling from a normal distribution at every time 
step. In order to ensure the approach to the correct long time limit of the equilibrium Boltzmann-Juttner distribution, 
we take
\beq
D=\gamma E T\label{eq.D}
\eeq
for a HQ with mass $m$ and energy $E=\sqrt{p^2+m^2}$ and adopt the post-point realization of 
the stochastic term~\cite{He:2013zua}. 
The HQ momentum and the EM field are boosted to the local rest frame 
of the fluid after which the Langevin updates are performed followed by the HQ momentum being reverted back to the lab frame. 
At the end of the Langevin evolution  \d and \dbar mesons are produced 
following the Peterson fragmentation of HQs~\cite{Peterson:1982ak}.

The Lorentz force ${\bf F}^{EM}$ is given by,
\beq
{\bf F}^{EM} = q\l {\bf E} + \l \frac{{\bf p}}{E}\times {\bf B}\r \r\label{eq.lorentz}
\eeq
where ${\bf E}$ and ${\bf B}$ are the electric and magnetic fields respectively induced by the protons in the colliding 
nuclei as well as the backreaction of the fireball with conductivity $\sigma$. The larger the value of $\sigma$ the  longer is 
the lifetime of the EM fields. The symmetry of the problem is such that the only relevant components for the computation 
of the directed flow are ${ B}_y$ and ${E}_x$. The calculation of the 
time dependent EM field follows Refs.~\cite{Tuchin:2013apa,Gursoy:2014aka}.
 We take only the contribution of
 the spectator protons to the EM fields and use a constant $\sigma$.  
The important thing to note is that the additional 
factor of ${ p}/E\sim0.3$ makes the magnetic force smaller compared to the electric force which makes the slope of the 
\d meson $v_1$ larger than \dbare. 

To study the dependence on the initial time  we  vary $\tau_0$ between 0.2 fm to 0.6 fm. Several lattice QCD computations 
suggest $\sigma\sim0.023$ fm$^{-1}$ around $2T_c$~\cite{Ding:2010ga,*Amato:2013naa,*Brandt:2012jc}. We vary $\sigma$ in the 
range  0.011 - 0.035 fm$^{-1}$. $D$ could be obtained from the scattering matrix formalism~\cite{vanHees:2004gq,
*Moore:2004tg,*Gubser:2006qh,*Alberico:2011zy,*Berrehrah:2014tva,*Scardina:2017ipo}. This also fixes $\gamma$ via 
Eq.~\ref{eq.D}. However, we adopt a data driven approach~\cite{Xu:2017obm, Chatterjee:2017ahy}. We work with a simple ansatz 
of $\gamma\propto T\l T/m\r^x$. In Ref.~\cite{Chatterjee:2017ahy}, it was shown 
that one gets a good qualitative description of the $p_T$ dependence of $R_{AA}$ and $v_2$ at mid-rapidity with a choice of 
$x$ between 0 to 0.5. Hence, we  implement these two extreme choices for $\gamma$, $\gamma\propto T$ (large drag) and 
$\gamma\propto T^{1.5}$ (small drag).

\begin{figure}
 \begin{center}
 \includegraphics[scale=0.44]{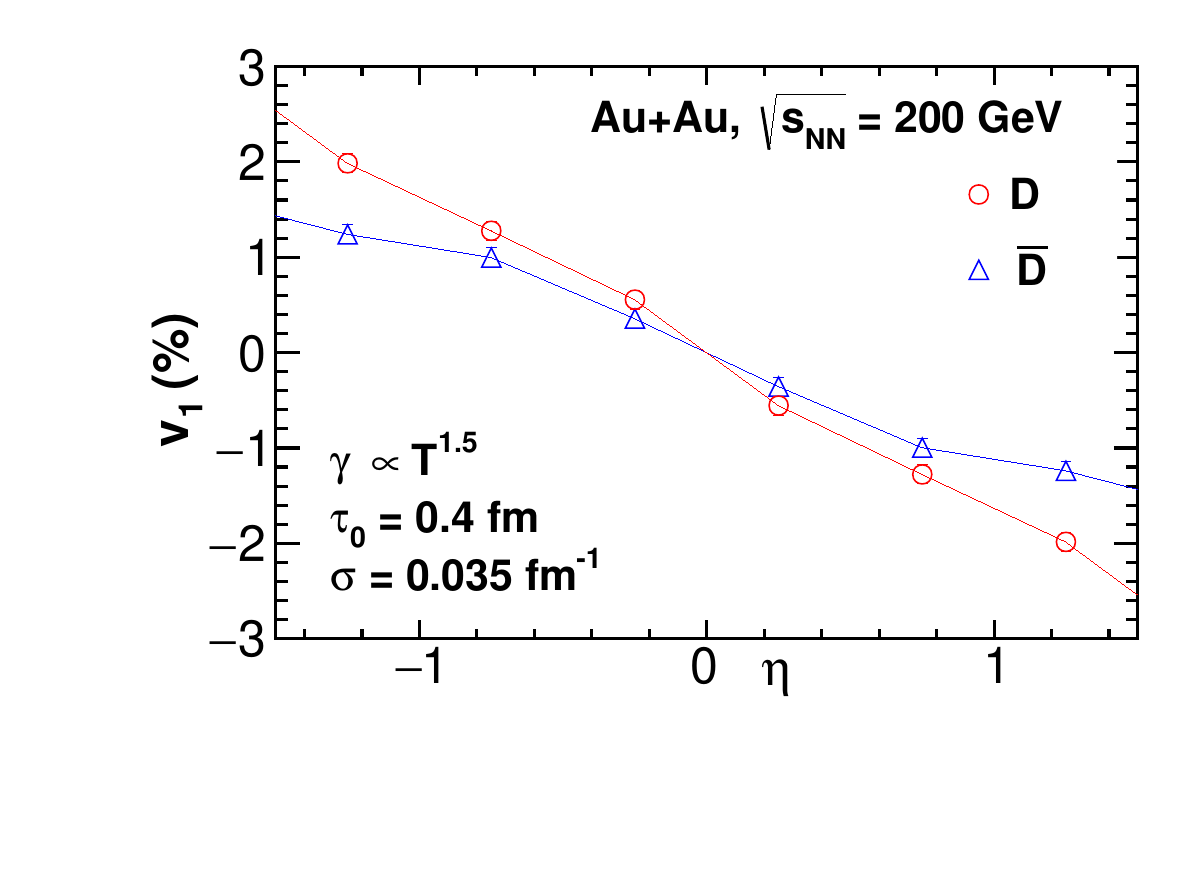}
 \caption{(Color online) The rapidity dependence of the directed flow coefficient $v_1$ for \d and \dbar mesons. The drag by the 
 tilted fireball on \d and \dbar, being charge independent, creates a rapidity-odd $v_1$ of same sign and 
 strength. The Lorentz force due to the EM field sourced by the protons in the colliding nuclei 
 results in opposite-sign contributions to the  rapidity-odd $v_1$ for \d and \dbar. }
 \label{fig.v1}
 \end{center}
\end{figure}

In Fig.~\ref{fig.v1} is shown the $v_1$ of \d and \dbar mesons resulting from the dynamics including  the combined influence of 
the drag by matter in the tilted fireball as well as by the Lorentz force. The
harmonic flow coefficient  $v_1$ is obtained as follows
\beq
v_1 = \la \cos\l \phi - \Psi_1\r\ra\label{eq.v1}
\eeq
where $\phi$ is the $D$ azimuthal angle, $\Psi_1$ is the  reaction plane of the event and $\la..\ra$ represents 
an ensemble average over realizations of the Langevin evolution. 
 In our calculation the reaction plane is well defined by the geometry of the event, in experiment it can be reconstructed from the 
 spectators. We use the same definition of the  reaction plane as in 
experimental analyzes \cite{Abelev:2008jga,Abelev:2013cva};  the nucleus flying in the positive $\eta$ direction is located at positive $x$. Note that it
 is the reverse of the orientation  used in Ref. \cite{Gursoy:2014aka}.

In terms of the average \vavg and difference 
\vdiff of \d and \dbar flows,
\beqa
 v_1^\text{avg} &=& \frac{1}{2}\l v_1\l D^0\r + v_1\l \overline{D^0}\r\r\label{eq.sumv1}\\
 v_1^{\text{diff}} &=& v_1\l D^0\r - v_1\l \bar{D^0}\r\label{eq.diffv1} 
\eeqa
the drag by the tilted fireball alone is expected to give rise to a non-zero $v_1^\text{avg}$ and no 
$v_1^{\text{diff}}$~\cite{Chatterjee:2017ahy}. 
On the other hand, the EM force alone can only give rise to $v_1^{\text{diff}}$ and zero $v_1^\text{avg}$~\cite{Das:2016cwd}. The combined 
effect due to both gives rise to both \vavg as well as \vdiffe. Fig.~\ref{fig.v1} suggests that the drag from the 
tilted source dominates and one obtains the same sign $v_1$ for both \d and \dbare. The  EM field gives rise to  a small charge splitting  of the directed flow
 $v_1^{\text{diff}}$. We now focus on the $v_1$ slope at mid-rapidity to make some quantitative statements 
on the dependence on  various model parameters like $\tau_0$, $\sigma$ and the collision energy $\sNN$.

\begin{figure}
 \begin{center}
  \includegraphics[scale=0.44]{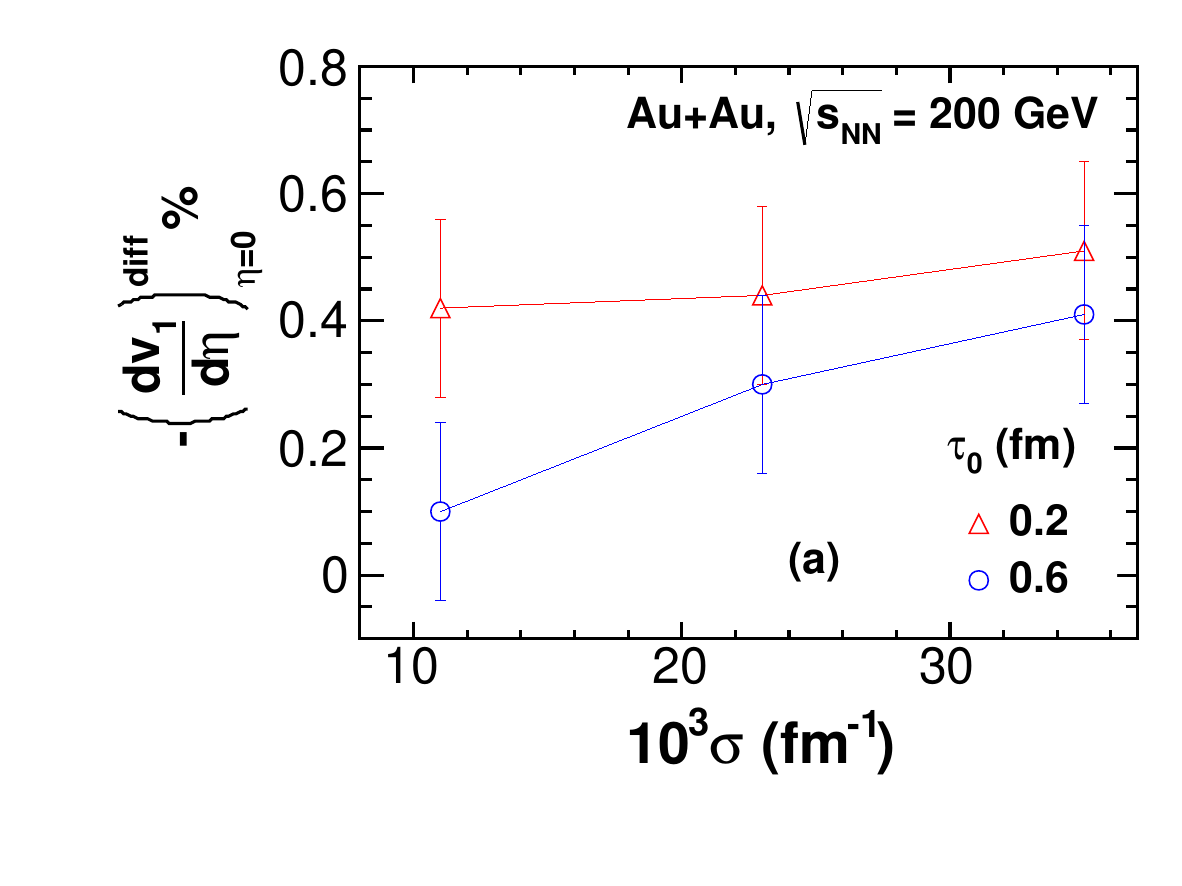}
 \includegraphics[scale=0.44]{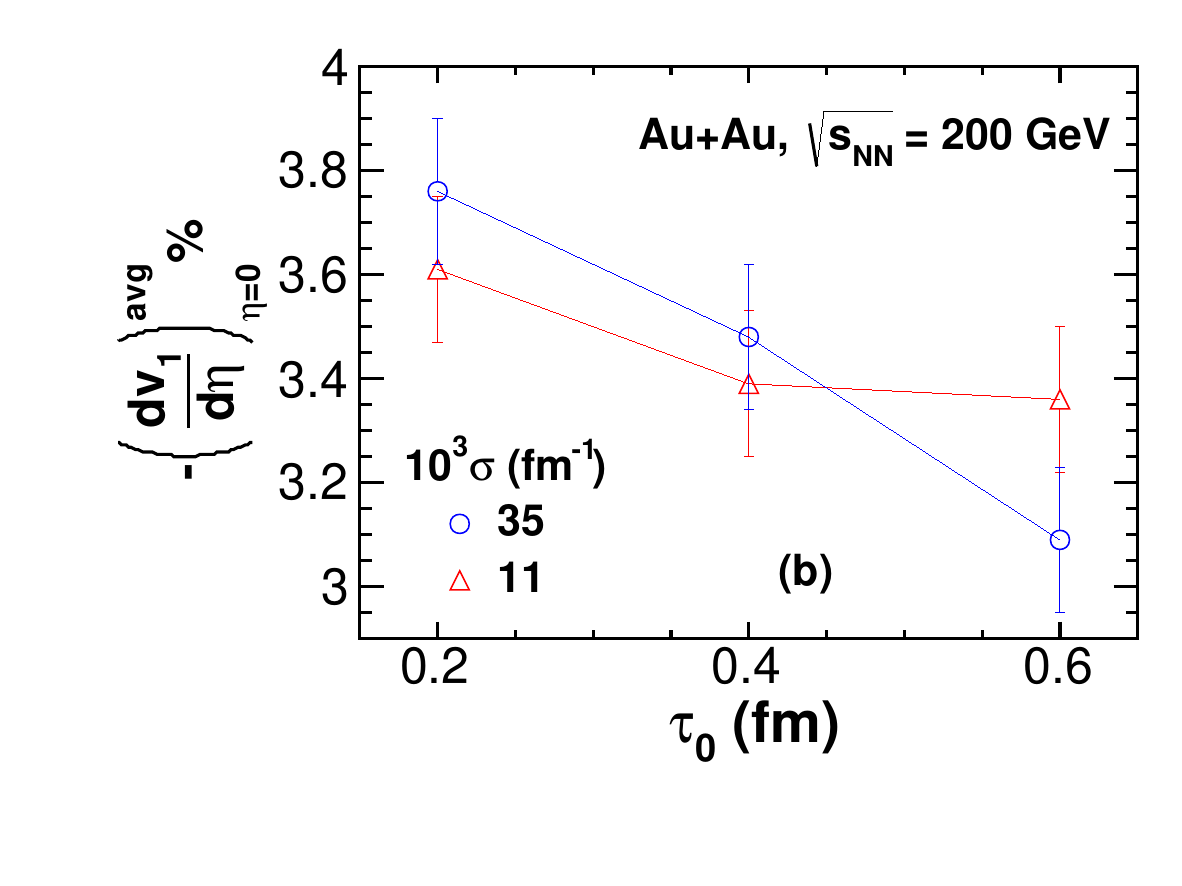}
 \caption{(Color online) (a) Charge splitting  $\l\frac{dv_1}{d\eta}\r^{\text{diff}}_{\eta=0}=
 \l\frac{dv_1}{d\eta}\r^{D}_{\eta=0}-\l\frac{dv_1}{d\eta}\r^{\overline{D}}_{\eta=0}$
 is plotted as a function of  the medium 
 conductivity $\sigma$. Results are  shown for two different initial times $\tau_0=0.2$ and 0.6 fm. (b) The mean 
 slope of  the $D$ meson directed flow , $\l\frac{dv_1}{d\eta}\r^{avg}_{\eta=0}=0.5\l\l\frac{dv_1}{d\eta}\r^{D}_{\eta=0}+
 \l\frac{dv_1}{d\eta}\r^{\overline{D}}_{\eta=0}\r$ is plotted as a function of the  initial time $\tau_0$. }
 \label{fig.tau0sigma}
 \end{center}
\end{figure}

The parameter dependence of  \d and \dbar directed flow 
in Au+Au collisions at $\sNN=200$~GeV is shown in Fig.~\ref{fig.tau0sigma}.
Fig.~\ref{fig.tau0sigma} (a) shows the $\eta$ slope of 
\vdiff with respect to variation in $\sigma$. As $\sigma$ is raised, the charge splitting of the slope  at  midrapidity  \svdiff 
raises by $400\%$ for $\tau_0=0.6$fm/c and by $25$\% for $\tau_0=0.2$fm/c. Thus, our results show the possibility to 
extract $\sigma$ of QCD matter with the observation of the charge splitting at midrapidity \svdiff. Also, a smaller $\tau_0$ consistently yields a larger \svdiff. 
This is because, with increasing $\sigma$ and/or decreasing $\tau_0$, the Lorentz force acts for a longer time on the HQs resulting in increasing \vdiffe.
At very early times ($\tau\lesssim0.2$ fm), the trend of the {\bf E} field is quite different~\cite{Das:2016cwd} which might lead to reduction of the charge splitting. This 
could be relevant within a framework that takes into account preequilibrium dynamics of the HQs.

Fig.~\ref{fig.tau0sigma} (b) shows the variation of \svavg 
with $\tau_0$. \svavg is around ten times larger than \svdiff. 
As $\tau_0$ is raised from $0.2$ to $0.6$ fm/c, \svavg falls by about $20\%$. 
The average directed flow of charm quarks  \svavg is built up in the early
 phase, lowering  $\tau_0$ increases the interaction of the HQ with the dense medium.
There is no significant  dependence of \svavg on the electric conductivity $\sigma$, as expected. \svavg is most sensitive to the choice of tilt angle in the initial state
which was studied in Ref.~\cite{Chatterjee:2017ahy}.

\begin{figure}
 \begin{center}
  \includegraphics[scale=0.44]{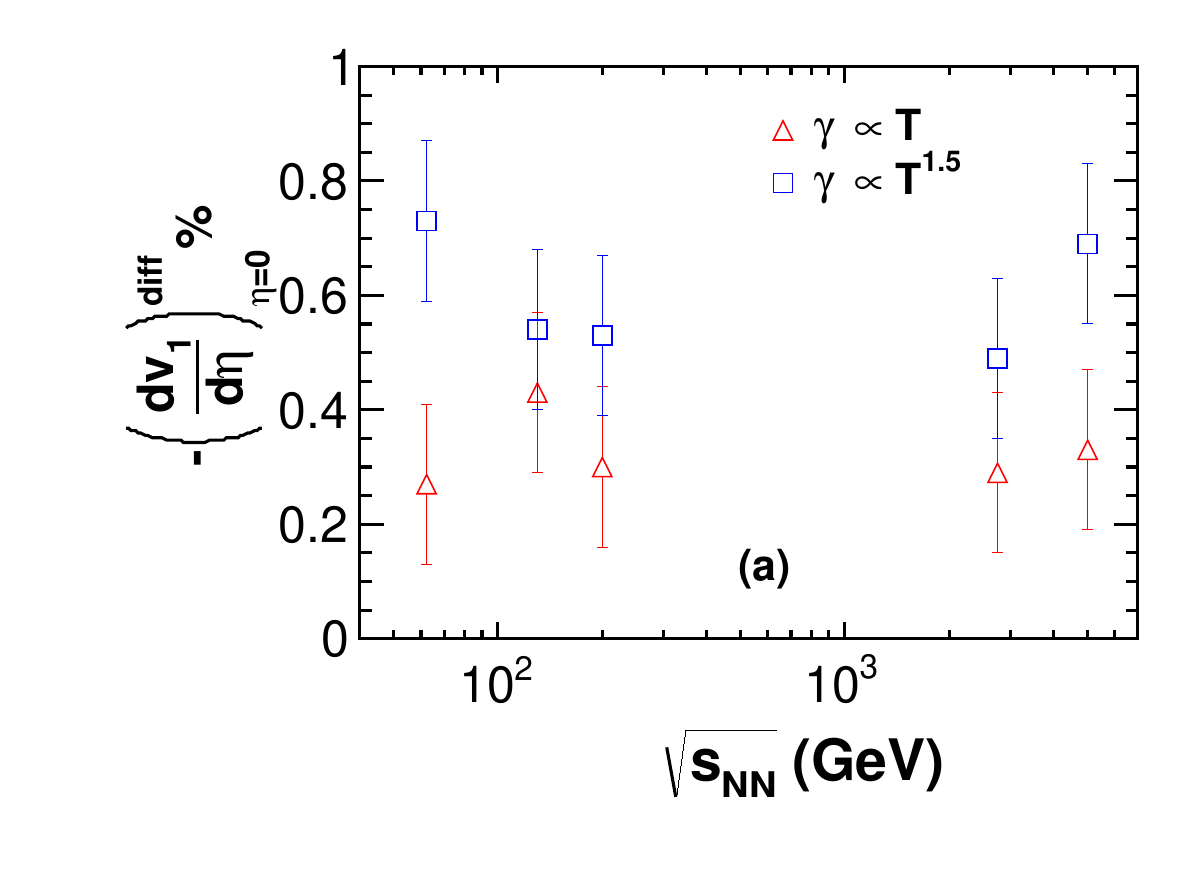} 
  \includegraphics[scale=0.44]{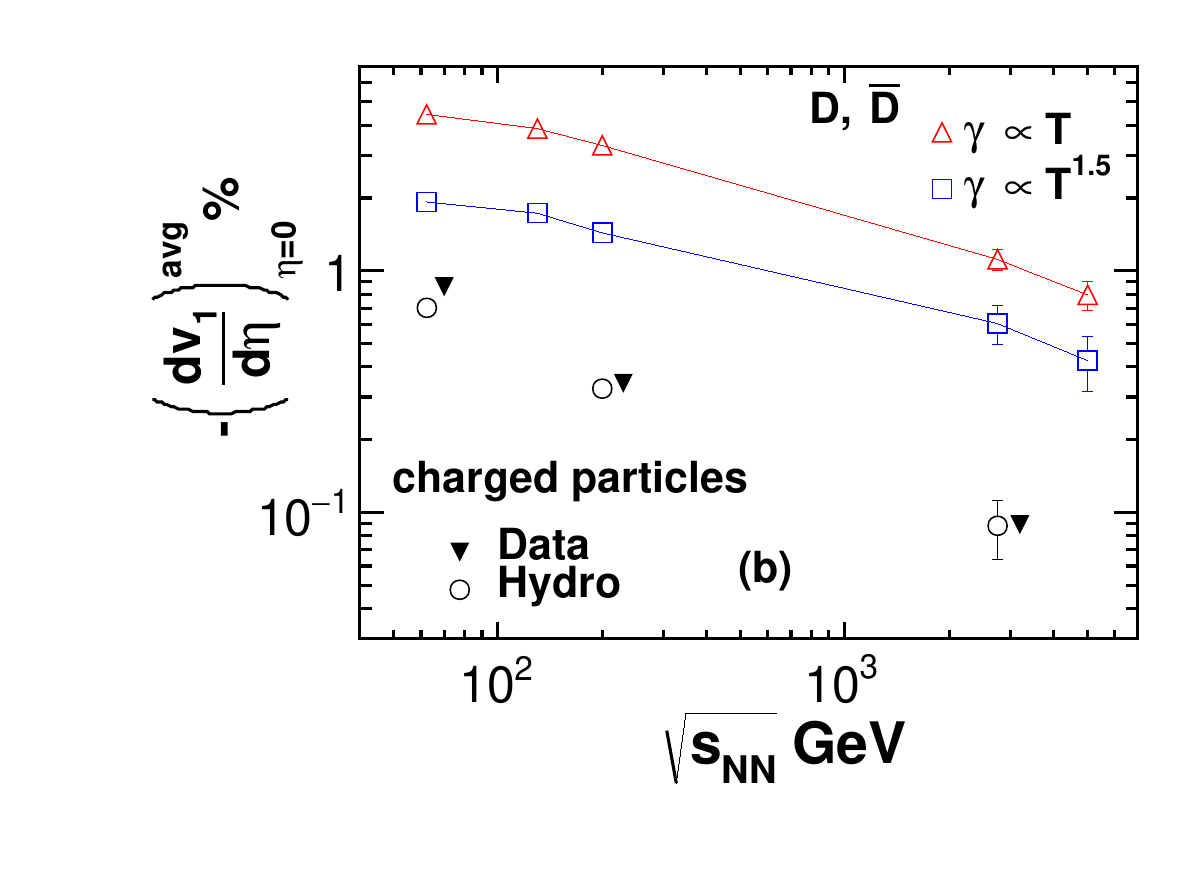}
 \caption{(Color online) (a) Charge splitting of the $D$ meson flow  $\l\frac{dv_1}{d\eta}\r^{\text{diff}}_{\eta=0}$ as a function of the collision energy $\sNN$. 
(b) Charge averaged directed flow  $\l\frac{dv_1}{d\eta}\r^{\text{avg}}_{\eta=0}$ 
as function of $\sNN$. The measured charged particle $v_1$ slope 
 at mid-rapidity~\cite{Abelev:2008jga,Abelev:2013cva} are denoted using full triangles; values obtained  in hydrodynamic calculations with suitably tuned parameters are also shown with open circles.
 All the results are  for $\sigma=0.023$ fm$^{-1}$ and $\tau_0=0.6$ fm.}
 \label{fig.rootS}
 \end{center}
\end{figure}

The dependence of the $D$ meson directed flow on the collision energy $\sNN$ is shown  for $\sigma=0.023$ fm$^{-1}$ and $\tau_0=0.6$ fm. 
 The dependence of the charge splitting of $D$ meson directed flow
 \svdiff on collision energy  is mostly flat (Fig. \ref{fig.rootS} (a)). In 
Fig.~\ref{fig.rootS} (b)  is plotted the slope of the average $v_1$ of \d and \dbare, \svavg versus
 $\sNN$. The experimental data for the measured charged particle $v_1$ slope 
is also shown for comparison~\cite{Abelev:2008jga,Abelev:2013cva}. This data serves to constrain the parameters of hydrodynamic calculation, and is reasonably well described by the model. On the other hand the energy  dependence of the  directed flow of $D$ mesons is a prediction.
Both  the bulk and the open charm $v_1$  decrease with increasing $\sNN$.  
Results for two choices of $\gamma$ is 
 shown to gauge the uncertainty in the 
prediction for different scenarios of the temperature dependence of the drag coefficient. 
The $\gamma\propto T$ (stronger drag) 
results show a larger \svavg and a smaller \svdiff while 
the $\gamma\propto T^{1.5}$ (weaker drag) results show a smaller \svavg and a larger \svdiff. This clearly reveals the 
interplay of drag by the expanding tilted source and the Lorentz force by the EM field; a stronger drag shifts the balance more in favor of the 
charge independent flow  by the tilted bulk resulting in a larger value of \svavg and a smaller value of \svdiff and vice versa for a 
weaker drag.

\begin{figure}
 \begin{center}
 \includegraphics[scale=0.44]{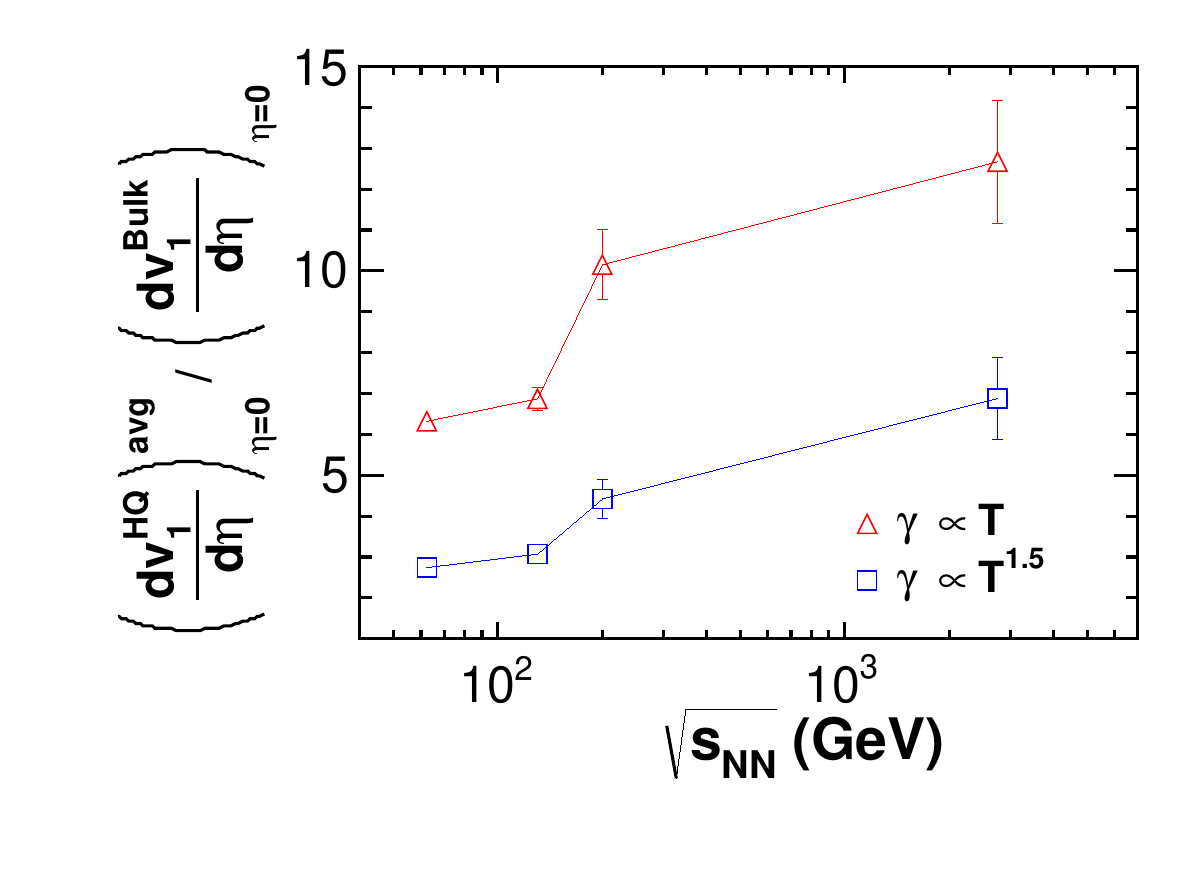}
 \caption{(Color online) The beam energy dependence of the ratio of the average heavy quark $v_1$ slope to that of the 
 charged particle $v_1$ is shown. Calculations are performed for $\sigma=0.023$ fm$^{-1}$ and $\tau_0=0.6$ fm and two choices of the temperature dependence of the HQ drag coefficient.}
 \label{fig.ratio}
 \end{center}
\end{figure}

Fig.~\ref{fig.ratio} shows the collision energy dependence of the ratio of \svavg for $D$ mesons  to the value for charged particles.
 While as seen 
in Fig.~\ref{fig.rootS} (b), both the numerator and denominator show a decreasing trend with $\sNN$ due to a reduction of the   tilt angle, we 
find that the decrease of the heavy flavor $v_1$ slope is smaller compared to that of the bulk resulting in the increasing trend for their 
ratio in Fig.~\ref{fig.ratio}. This increasing ratio stems from the fact that with increasing $\sNN$, the fireball is  denser  that calls for stronger drag by the bulk on the HQ and hence a relatively larger $v_1$ of the final $D$ mesons
at LHC energies.

HQs serve as good probes of the initial condition in heavy-ion collisions by virtue of being produced only in the initial 
state. We study the combined effects of two initial phenomena on the flow pattern of HQs, the drag from the expanding  tilted 
 fireball and the large EM field in the early stage of the collision.
 Both of these give rise to HQ directed flow. While the charge independent drag   by the matter in the fireball 
gives rise to the same  $v_1$ for \d and \dbare, the charge dependent Lorentz force by the EM fields gives rise to unlike sign contribution to 
$v_1$ of \d and \dbare. We find that the HQ drag contribution  dominates 
resulting in same sign $v_1$ for both \d and \dbar with their average $v_1$ being 10 times larger than their difference. The 
sensitivity of the charge splitting and average $v_1$ of \d and \dbar on the model parameters like $\sigma$, $\tau_0$ and $\gamma$ is studied.
 A smaller $\tau_0$ and/or larger $\sigma$ lengthens the time over which the Lorentz force acts on the HQs. This results in 
larger charge splitting \svdiffe. Also a smaller $\gamma$ reduces the
 opacity of the tilted source and hence again raises \svdiffe. A smaller
 initial time $\tau_0$ for the formation of the fireball means that HQ feel the drag of the opaque, dense matter in fireball for a longer time.  

The energy dependence  of the observed phenomena stems from three main effects. A decrease of the fireball tilt with energy, resulting in a decrease of the directed flow for both charged particles and $D$ mesons. An increase of the fireball temperature, which makes the fireball more opaque to HQ. 
This effect counterbalances to some extent the first one for $D$ mesons.
 When going from RHIC to LHC the average directed flow of $D$ mesons
  is reduced less than that of charged particles. The third effect, taken into account in our calculation, is the energy dependence of the dynamics of the EM fields in the collision. The resulting directed flow splitting for \d and \dbar  
is found to have an  
almost a flat dependence on $\sNN$.

This work is the first study of the combined effect of the initial tilt of 
the fireball and the large initial EM fields on the directed flow of \d and \dbar mesons. 
There are several systematic effects which could be investigated to quantify their influence 
on the numerical estimates that are presented here. Apart from fragmentation, harmonization could 
also take place via quark recombination or coalescence~\cite{Greco:2003vf,*vanHees:2005wb,*Cao:2013ita,
*Song:2015ykw,*Plumari:2017ntm}. Since, the light flavor $v_1$ is 
much smaller than heavy quarks, we expect small influence on the final state heavy flavor $v_1$.
The effect of the hadronic rescattering phase post chemical freezeout on the \d and \dbar $v_1$ 
could also be studied in the future~\cite{Das:2015ana,Cao:2015hia}.

Note added: After the completion of this work preliminary experimental data on the directed flow 
of charm mesons appeared. The results from the STAR Collaboration for Au-Au collisions 
at $\sqrt{s_{NN}}=200$~MeV \cite{Singha:2018cdj} is in qualitative agreement with our predictions. 
The directed flow of charm mesons is large and the charge splitting of the charm meson flow is 
small (if any). Our calculation predicts a weak energy dependency of the charge splitting of the 
charm meson directed flow. Preliminary results of the ALICE Collaboration for Pb-Pb collisions 
at $\sqrt{s_{NN}}=5.02$~TeV \cite{Grosa:2018zix} indicate an order of magnitude larger charge 
splitting of the directed flow. In our calculation such a strong change with energy cannot be 
explained using a mechanism based on the diffusion of heavy quarks in electromagnetic fields 
using similar  model parameters at RHIC and LHC energies.

This research  is supported by the AGH UST statutory tasks
 No.~11.11.220.01/1 within subsidy of the Polish Ministry of 
Science and Higher Education and the National Science Centre 
Grant No. 2015/17/B/ST2/00101.

\bibliographystyle{apsrev4-1-nohep}
\bibliography{HQTiltEM}

\begin{thebibliography}{10}%
\makeatletter
\providecommand \@ifxundefined [1]{%
 \ifx #1\undefined \expandafter \@firstoftwo
 \else \expandafter \@secondoftwo
\fi
}%
\providecommand \@ifnum [1]{%
 \ifnum #1\expandafter \@firstoftwo
 \else \expandafter \@secondoftwo
\fi
}%
\providecommand \enquote [1]{``#1''}%
\providecommand \bibnamefont  [1]{#1}%
\providecommand \bibfnamefont [1]{#1}%
\providecommand \citenamefont [1]{#1}%
\providecommand\href[0]{\@sanitize\@href}%
\providecommand\@href[1]{\endgroup\@@startlink{#1}\endgroup\@@href}%
\providecommand\@@href[1]{#1\@@endlink}%
\providecommand \@sanitize [0]{\begingroup\catcode`\&12\catcode`\#12\relax}%
\@ifxundefined \pdfoutput {\@firstoftwo}{%
 \@ifnum{\z@=\pdfoutput}{\@firstoftwo}{\@secondoftwo}%
}{%
 \providecommand\@@startlink[1]{\leavevmode\special{html:<a href="#1">}}%
 \providecommand\@@endlink[0]{\special{html:</a>}}%
}{%
 \providecommand\@@startlink[1]{%
  \leavevmode
  \pdfstartlink
   attr{/Border[0 0 1 ]/H/I/C[0 1 1]}%
   user{/Subtype/Link/A<</Type/Action/S/URI/URI(#1)>>}%
  \relax
 }%
 \providecommand\@@endlink[0]{\pdfendlink}%
}%
\providecommand \url  [0]{\begingroup\@sanitize \@url }%
\providecommand \@url [1]{\endgroup\@href {#1}{\urlprefix}}%
\providecommand \urlprefix [0]{URL }%
\providecommand \Eprint[0]{\href }%
\@ifxundefined \urlstyle {%
  \providecommand \doi [1]{doi:\discretionary{}{}{}#1}%
}{%
  \providecommand \doi [0]{doi:\discretionary{}{}{}\begingroup
  \urlstyle{rm}\Url }%
}%
\providecommand \doibase [0]{http://dx.doi.org/}%
\providecommand \Doi[1]{\href{\doibase#1}}%
\providecommand \bibAnnote [3]{%
  \BibitemShut{#1}%
  \begin{quotation}\noindent
    \textsc{Key:}\ #2\\\textsc{Annotation:}\ #3%
  \end{quotation}%
}%
\providecommand \bibAnnoteFile [2]{%
  \IfFileExists{#2}{\bibAnnote {#1} {#2} {\input{#2}}}{}%
}%
\providecommand \typeout [0]{\immediate \write \m@ne }%
\providecommand \selectlanguage [0]{\@gobble}%
\providecommand \bibinfo [0]{\@secondoftwo}%
\providecommand \bibfield [0]{\@secondoftwo}%
\providecommand \translation [1]{[#1]}%
\providecommand \BibitemOpen[0]{}%
\providecommand \bibitemStop [0]{}%
\providecommand \bibitemNoStop [0]{.\EOS\space}%
\providecommand \EOS [0]{\spacefactor3000\relax}%
\providecommand \BibitemShut [1]{\csname bibitem#1\endcsname}%
\bibitem{Ollitrault:2010tn}%
  \BibitemOpen
  \bibfield{author}{%
  \bibinfo {author} {\bibfnamefont{J.-Y.}\ \bibnamefont{Ollitrault}},\ }%
  \bibfield{journal}{%
  \Doi{10.1088/1742-6596/312/1/012002}{\bibinfo {journal} {J. Phys. Conf.
  Ser.}}\ }%
  \textbf{\bibinfo {volume} {312}},\ \bibinfo {pages} {012002} (\bibinfo {year}
  {2011})%
  \bibAnnoteFile{NoStop}{Ollitrault:2010tn}%
\bibitem{Heinz:2013th}%
  \BibitemOpen
  \bibfield{author}{%
  \bibinfo {author} {\bibfnamefont{U.}~\bibnamefont{Heinz}}\ and\ \bibinfo
  {author} {\bibfnamefont{R.}~\bibnamefont{Snellings}},\ }%
  \bibfield{journal}{%
  \Doi{10.1146/annurev-nucl-102212-170540}{\bibinfo {journal} {Ann. Rev. Nucl.
  Part. Sci.}}\ }%
  \textbf{\bibinfo {volume} {63}},\ \bibinfo {pages} {123} (\bibinfo {year}
  {2013})%
  \bibAnnoteFile{NoStop}{Heinz:2013th}%
\bibitem{Gale:2013da}%
  \BibitemOpen
  \bibfield{author}{%
  \bibinfo {author} {\bibfnamefont{C.}~\bibnamefont{Gale}}, \bibinfo {author}
  {\bibfnamefont{S.}~\bibnamefont{Jeon}},\ and\ \bibinfo {author}
  {\bibfnamefont{B.}~\bibnamefont{Schenke}},\ }%
  \bibfield{journal}{%
  \Doi{10.1142/S0217751X13400113}{\bibinfo {journal} {Int. J. Mod. Phys.}}\ }%
  \textbf{\bibinfo {volume} {A28}},\ \bibinfo {pages} {1340011} (\bibinfo
  {year} {2013})%
  \bibAnnoteFile{NoStop}{Gale:2013da}%
\bibitem{Gupta:2003zh}%
  \BibitemOpen
  \bibfield{author}{%
  \bibinfo {author} {\bibfnamefont{S.}~\bibnamefont{Gupta}},\ }%
  \bibfield{journal}{%
  \Doi{10.1016/j.physletb.2004.05.079}{\bibinfo {journal} {Phys. Lett.}}\ }%
  \textbf{\bibinfo {volume} {B597}},\ \bibinfo {pages} {57} (\bibinfo {year}
  {2004})%
  \bibAnnoteFile{NoStop}{Gupta:2003zh}%
\bibitem{Hirono:2012rt}%
  \BibitemOpen
  \bibfield{author}{%
  \bibinfo {author} {\bibfnamefont{Y.}~\bibnamefont{Hirono}}, \bibinfo {author}
  {\bibfnamefont{M.}~\bibnamefont{Hongo}},\ and\ \bibinfo {author}
  {\bibfnamefont{T.}~\bibnamefont{Hirano}},\ }%
  \bibfield{journal}{%
  \Doi{10.1103/PhysRevC.90.021903}{\bibinfo {journal} {Phys. Rev.}}\ }%
  \textbf{\bibinfo {volume} {C90}},\ \bibinfo {pages} {021903} (\bibinfo {year}
  {2014})%
  \bibAnnoteFile{NoStop}{Hirono:2012rt}%
\bibitem{Cassing:2013iz}%
  \BibitemOpen
  \bibfield{author}{%
  \bibinfo {author} {\bibfnamefont{W.}~\bibnamefont{Cassing}}, \bibinfo
  {author} {\bibfnamefont{O.}~\bibnamefont{Linnyk}}, \bibinfo {author}
  {\bibfnamefont{T.}~\bibnamefont{Steinert}},\ and\ \bibinfo {author}
  {\bibfnamefont{V.}~\bibnamefont{Ozvenchuk}},\ }%
  \bibfield{journal}{%
  \Doi{10.1103/PhysRevLett.110.182301}{\bibinfo {journal} {Phys. Rev. Lett.}}\
  }%
  \textbf{\bibinfo {volume} {110}},\ \bibinfo {pages} {182301} (\bibinfo {year}
  {2013})%
  \bibAnnoteFile{NoStop}{Cassing:2013iz}%
\bibitem{Yin:2013kya}%
  \BibitemOpen
  \bibfield{author}{%
  \bibinfo {author} {\bibfnamefont{Y.}~\bibnamefont{Yin}},\ }%
  \bibfield{journal}{%
  \Doi{10.1103/PhysRevC.90.044903}{\bibinfo {journal} {Phys. Rev.}}\ }%
  \textbf{\bibinfo {volume} {C90}},\ \bibinfo {pages} {044903} (\bibinfo {year}
  {2014})%
  \bibAnnoteFile{NoStop}{Yin:2013kya}%
\bibitem{Finazzo:2013efa}%
  \BibitemOpen
  \bibfield{author}{%
  \bibinfo {author} {\bibfnamefont{S.~I.}\ \bibnamefont{Finazzo}}\ and\
  \bibinfo {author} {\bibfnamefont{J.}~\bibnamefont{Noronha}},\ }%
  \bibfield{journal}{%
  \Doi{10.1103/PhysRevD.89.106008}{\bibinfo {journal} {Phys. Rev.}}\ }%
  \textbf{\bibinfo {volume} {D89}},\ \bibinfo {pages} {106008} (\bibinfo {year}
  {2014})%
  \bibAnnoteFile{NoStop}{Finazzo:2013efa}%
\bibitem{Puglisi:2014sha}%
  \BibitemOpen
  \bibfield{author}{%
  \bibinfo {author} {\bibfnamefont{A.}~\bibnamefont{Puglisi}}, \bibinfo
  {author} {\bibfnamefont{S.}~\bibnamefont{Plumari}},\ and\ \bibinfo {author}
  {\bibfnamefont{V.}~\bibnamefont{Greco}},\ }%
  \bibfield{journal}{%
  \Doi{10.1103/PhysRevD.90.114009}{\bibinfo {journal} {Phys. Rev.}}\ }%
  \textbf{\bibinfo {volume} {D90}},\ \bibinfo {pages} {114009} (\bibinfo {year}
  {2014})%
  \bibAnnoteFile{NoStop}{Puglisi:2014sha}%
\bibitem{Greif:2014oia}%
  \BibitemOpen
  \bibfield{author}{%
  \bibinfo {author} {\bibfnamefont{M.}~\bibnamefont{Greif}}, \bibinfo {author}
  {\bibfnamefont{I.}~\bibnamefont{Bouras}}, \bibinfo {author}
  {\bibfnamefont{C.}~\bibnamefont{Greiner}},\ and\ \bibinfo {author}
  {\bibfnamefont{Z.}~\bibnamefont{Xu}},\ }%
  \bibfield{journal}{%
  \Doi{10.1103/PhysRevD.90.094014}{\bibinfo {journal} {Phys. Rev.}}\ }%
  \textbf{\bibinfo {volume} {D90}},\ \bibinfo {pages} {094014} (\bibinfo {year}
  {2014})%
  \bibAnnoteFile{NoStop}{Greif:2014oia}%
\bibitem{Srivastava:2015via}%
  \BibitemOpen
  \bibfield{author}{%
  \bibinfo {author} {\bibfnamefont{P.~K.}\ \bibnamefont{Srivastava}}, \bibinfo
  {author} {\bibfnamefont{L.}~\bibnamefont{Thakur}},\ and\ \bibinfo {author}
  {\bibfnamefont{B.~K.}\ \bibnamefont{Patra}},\ }%
  \bibfield{journal}{%
  \Doi{10.1103/PhysRevC.91.044903}{\bibinfo {journal} {Phys. Rev.}}\ }%
  \textbf{\bibinfo {volume} {C91}},\ \bibinfo {pages} {044903} (\bibinfo {year}
  {2015})%
  \bibAnnoteFile{NoStop}{Srivastava:2015via}%
\bibitem{Ghosh:2016yvt}%
  \BibitemOpen
  \bibfield{author}{%
  \bibinfo {author} {\bibfnamefont{S.}~\bibnamefont{Ghosh}},\ }%
  \bibfield{journal}{%
  \Doi{10.1103/PhysRevD.95.036018}{\bibinfo {journal} {Phys. Rev.}}\ }%
  \textbf{\bibinfo {volume} {D95}},\ \bibinfo {pages} {036018} (\bibinfo {year}
  {2017})%
  \bibAnnoteFile{NoStop}{Ghosh:2016yvt}%
\bibitem{Hattori:2016cnt}%
  \BibitemOpen
  \bibfield{author}{%
  \bibinfo {author} {\bibfnamefont{K.}~\bibnamefont{Hattori}}\ and\ \bibinfo
  {author} {\bibfnamefont{D.}~\bibnamefont{Satow}},\ }%
  \bibfield{journal}{%
  \Doi{10.1103/PhysRevD.94.114032}{\bibinfo {journal} {Phys. Rev.}}\ }%
  \textbf{\bibinfo {volume} {D94}},\ \bibinfo {pages} {114032} (\bibinfo {year}
  {2016})%
  \bibAnnoteFile{NoStop}{Hattori:2016cnt}%
\bibitem{Feng:2017tsh}%
  \BibitemOpen
  \bibfield{author}{%
  \bibinfo {author} {\bibfnamefont{B.}~\bibnamefont{Feng}},\ }%
  \bibfield{journal}{%
  \Doi{10.1103/PhysRevD.96.036009}{\bibinfo {journal} {Phys. Rev.}}\ }%
  \textbf{\bibinfo {volume} {D96}},\ \bibinfo {pages} {036009} (\bibinfo {year}
  {2017})%
  \bibAnnoteFile{NoStop}{Feng:2017tsh}%
\bibitem{Thakur:2017hfc}%
  \BibitemOpen
  \bibfield{author}{%
  \bibinfo {author} {\bibfnamefont{L.}~\bibnamefont{Thakur}}, \bibinfo {author}
  {\bibfnamefont{P.~K.}\ \bibnamefont{Srivastava}}, \bibinfo {author}
  {\bibfnamefont{G.~P.}\ \bibnamefont{Kadam}}, \bibinfo {author}
  {\bibfnamefont{M.}~\bibnamefont{George}},\ and\ \bibinfo {author}
  {\bibfnamefont{H.}~\bibnamefont{Mishra}},\ }%
  \bibfield{journal}{%
  \Doi{10.1103/PhysRevD.95.096009}{\bibinfo {journal} {Phys. Rev.}}\ }%
  \textbf{\bibinfo {volume} {D95}},\ \bibinfo {pages} {096009} (\bibinfo {year}
  {2017})%
  \bibAnnoteFile{NoStop}{Thakur:2017hfc}%
\bibitem{Mitra:2017sjo}%
  \BibitemOpen
  \bibfield{author}{%
  \bibinfo {author} {\bibfnamefont{S.}~\bibnamefont{Mitra}}\ and\ \bibinfo
  {author} {\bibfnamefont{V.}~\bibnamefont{Chandra}},\ }%
  \bibfield{journal}{%
  \Doi{10.1103/PhysRevD.96.094003}{\bibinfo {journal} {Phys. Rev.}}\ }%
  \textbf{\bibinfo {volume} {D96}},\ \bibinfo {pages} {094003} (\bibinfo {year}
  {2017})%
  \bibAnnoteFile{NoStop}{Mitra:2017sjo}%
\bibitem{Ghosh:2018kst}%
  \BibitemOpen
  \bibfield{author}{%
  \bibinfo {author} {\bibfnamefont{S.}~\bibnamefont{Ghosh}}, \bibinfo {author}
  {\bibfnamefont{S.}~\bibnamefont{Mitra}},\ and\ \bibinfo {author}
  {\bibfnamefont{S.}~\bibnamefont{Sarkar}},\ }%
  \bibfield{journal}{%
  \Doi{10.1016/j.nuclphysa.2017.10.008}{\bibinfo {journal} {Nucl. Phys.}}\ }%
  \textbf{\bibinfo {volume} {A969}},\ \bibinfo {pages} {237} (\bibinfo {year}
  {2018})%
  \bibAnnoteFile{NoStop}{Ghosh:2018kst}%
\bibitem{Ding:2010ga}%
  \BibitemOpen
  \bibfield{author}{%
  \bibinfo {author} {\bibfnamefont{H.~T.}\ \bibnamefont{Ding}}, \bibinfo
  {author} {\bibfnamefont{A.}~\bibnamefont{Francis}}, \bibinfo {author}
  {\bibfnamefont{O.}~\bibnamefont{Kaczmarek}}, \bibinfo {author}
  {\bibfnamefont{F.}~\bibnamefont{Karsch}}, \bibinfo {author}
  {\bibfnamefont{E.}~\bibnamefont{Laermann}},\ and\ \bibinfo {author}
  {\bibfnamefont{W.}~\bibnamefont{Soeldner}},\ }%
  \bibfield{journal}{%
  \Doi{10.1103/PhysRevD.83.034504}{\bibinfo {journal} {Phys. Rev.}}\ }%
  \textbf{\bibinfo {volume} {D83}},\ \bibinfo {pages} {034504} (\bibinfo {year}
  {2011})%
  \bibAnnoteFile{NoStop}{Ding:2010ga}%
\bibitem{Amato:2013naa}%
  \BibitemOpen
  \bibfield{author}{%
  \bibinfo {author} {\bibfnamefont{A.}~\bibnamefont{Amato}}, \bibinfo {author}
  {\bibfnamefont{G.}~\bibnamefont{Aarts}}, \bibinfo {author}
  {\bibfnamefont{C.}~\bibnamefont{Allton}}, \bibinfo {author}
  {\bibfnamefont{P.}~\bibnamefont{Giudice}}, \bibinfo {author}
  {\bibfnamefont{S.}~\bibnamefont{Hands}},\ and\ \bibinfo {author}
  {\bibfnamefont{J.-I.}\ \bibnamefont{Skullerud}},\ }%
  \bibfield{journal}{%
  \Doi{10.1103/PhysRevLett.111.172001}{\bibinfo {journal} {Phys. Rev. Lett.}}\
  }%
  \textbf{\bibinfo {volume} {111}},\ \bibinfo {pages} {172001} (\bibinfo {year}
  {2013})%
  \bibAnnoteFile{NoStop}{Amato:2013naa}%
\bibitem{Brandt:2012jc}%
  \BibitemOpen
  \bibfield{author}{%
  \bibinfo {author} {\bibfnamefont{B.~B.}\ \bibnamefont{Brandt}}, \bibinfo
  {author} {\bibfnamefont{A.}~\bibnamefont{Francis}}, \bibinfo {author}
  {\bibfnamefont{H.~B.}\ \bibnamefont{Meyer}},\ and\ \bibinfo {author}
  {\bibfnamefont{H.}~\bibnamefont{Wittig}},\ }%
  \bibfield{journal}{%
  \Doi{10.1007/JHEP03(2013)100}{\bibinfo {journal} {JHEP}}\ }%
  \textbf{\bibinfo {volume} {03}},\ \bibinfo {pages} {100} (\bibinfo {year}
  {2013})%
  \bibAnnoteFile{NoStop}{Brandt:2012jc}%
\bibitem{Kharzeev:2007jp}%
  \BibitemOpen
  \bibfield{author}{%
  \bibinfo {author} {\bibfnamefont{D.~E.}\ \bibnamefont{Kharzeev}}, \bibinfo
  {author} {\bibfnamefont{L.~D.}\ \bibnamefont{McLerran}},\ and\ \bibinfo
  {author} {\bibfnamefont{H.~J.}\ \bibnamefont{Warringa}},\ }%
  \bibfield{journal}{%
  \Doi{10.1016/j.nuclphysa.2008.02.298}{\bibinfo {journal} {Nucl. Phys.}}\ }%
  \textbf{\bibinfo {volume} {A803}},\ \bibinfo {pages} {227} (\bibinfo {year}
  {2008})%
  \bibAnnoteFile{NoStop}{Kharzeev:2007jp}%
\bibitem{Tuchin:2010vs}%
  \BibitemOpen
  \bibfield{author}{%
  \bibinfo {author} {\bibfnamefont{K.}~\bibnamefont{Tuchin}},\ }%
  \bibfield{journal}{%
  \Doi{10.1103/PhysRevC.83.039903, 10.1103/PhysRevC.82.034904}{\bibinfo
  {journal} {Phys. Rev.}}\ }%
  \textbf{\bibinfo {volume} {C82}},\ \bibinfo {pages} {034904} (\bibinfo {year}
  {2010}),\ \bibinfo {note} {[Erratum: Phys. Rev.C83,039903(2011)]}%
  \bibAnnoteFile{NoStop}{Tuchin:2010vs}%
\bibitem{Bzdak:2011yy}%
  \BibitemOpen
  \bibfield{author}{%
  \bibinfo {author} {\bibfnamefont{A.}~\bibnamefont{Bzdak}}\ and\ \bibinfo
  {author} {\bibfnamefont{V.}~\bibnamefont{Skokov}},\ }%
  \bibfield{journal}{%
  \Doi{10.1016/j.physletb.2012.02.065}{\bibinfo {journal} {Phys. Lett.}}\ }%
  \textbf{\bibinfo {volume} {B710}},\ \bibinfo {pages} {171} (\bibinfo {year}
  {2012})%
  \bibAnnoteFile{NoStop}{Bzdak:2011yy}%
\bibitem{Voronyuk:2011jd}%
  \BibitemOpen
  \bibfield{author}{%
  \bibinfo {author} {\bibfnamefont{V.}~\bibnamefont{Voronyuk}}, \bibinfo
  {author} {\bibfnamefont{V.~D.}\ \bibnamefont{Toneev}}, \bibinfo {author}
  {\bibfnamefont{W.}~\bibnamefont{Cassing}}, \bibinfo {author}
  {\bibfnamefont{E.~L.}\ \bibnamefont{Bratkovskaya}}, \bibinfo {author}
  {\bibfnamefont{V.~P.}\ \bibnamefont{Konchakovski}},\ and\ \bibinfo {author}
  {\bibfnamefont{S.~A.}\ \bibnamefont{Voloshin}},\ }%
  \bibfield{journal}{%
  \Doi{10.1103/PhysRevC.83.054911}{\bibinfo {journal} {Phys. Rev.}}\ }%
  \textbf{\bibinfo {volume} {C83}},\ \bibinfo {pages} {054911} (\bibinfo {year}
  {2011})%
  \bibAnnoteFile{NoStop}{Voronyuk:2011jd}%
\bibitem{Deng:2012pc}%
  \BibitemOpen
  \bibfield{author}{%
  \bibinfo {author} {\bibfnamefont{W.-T.}\ \bibnamefont{Deng}}\ and\ \bibinfo
  {author} {\bibfnamefont{X.-G.}\ \bibnamefont{Huang}},\ }%
  \bibfield{journal}{%
  \Doi{10.1103/PhysRevC.85.044907}{\bibinfo {journal} {Phys. Rev.}}\ }%
  \textbf{\bibinfo {volume} {C85}},\ \bibinfo {pages} {044907} (\bibinfo {year}
  {2012})%
  \bibAnnoteFile{NoStop}{Deng:2012pc}%
\bibitem{Fukushima:2008xe}%
  \BibitemOpen
  \bibfield{author}{%
  \bibinfo {author} {\bibfnamefont{K.}~\bibnamefont{Fukushima}}, \bibinfo
  {author} {\bibfnamefont{D.~E.}\ \bibnamefont{Kharzeev}},\ and\ \bibinfo
  {author} {\bibfnamefont{H.~J.}\ \bibnamefont{Warringa}},\ }%
  \bibfield{journal}{%
  \Doi{10.1103/PhysRevD.78.074033}{\bibinfo {journal} {Phys. Rev.}}\ }%
  \textbf{\bibinfo {volume} {D78}},\ \bibinfo {pages} {074033} (\bibinfo {year}
  {2008})%
  \bibAnnoteFile{NoStop}{Fukushima:2008xe}%
\bibitem{Rapp:2009my}%
  \BibitemOpen
  \bibfield{author}{%
  \bibinfo {author} {\bibfnamefont{R.}~\bibnamefont{Rapp}}\ and\ \bibinfo
  {author} {\bibfnamefont{H.}~\bibnamefont{van Hees}}\ }%
  (\bibinfo {year} {2010})\ pp.\ \bibinfo {pages} {111--206},\
  \Eprint{http://arxiv.org/abs/0903.1096}{arXiv:0903.1096 [hep-ph]}%
  \bibAnnoteFile{NoStop}{Rapp:2009my}%
\bibitem{Andronic:2015wma}%
  \BibitemOpen
  \bibfield{author}{%
  \bibinfo {author} {\bibfnamefont{A.}~\bibnamefont{Andronic}} \emph{et~al.},\
  }%
  \bibfield{journal}{%
  \Doi{10.1140/epjc/s10052-015-3819-5}{\bibinfo {journal} {Eur. Phys. J.}}\ }%
  \textbf{\bibinfo {volume} {C76}},\ \bibinfo {pages} {107} (\bibinfo {year}
  {2016})%
  \bibAnnoteFile{NoStop}{Andronic:2015wma}%
\bibitem{Aarts:2016hap}%
  \BibitemOpen
  \bibfield{author}{%
  \bibinfo {author} {\bibfnamefont{G.}~\bibnamefont{Aarts}} \emph{et~al.},\ }%
  \bibfield{journal}{%
  \Doi{10.1140/epja/i2017-12282-9}{\bibinfo {journal} {Eur. Phys. J.}}\ }%
  \textbf{\bibinfo {volume} {A53}},\ \bibinfo {pages} {93} (\bibinfo {year}
  {2017})%
  \bibAnnoteFile{NoStop}{Aarts:2016hap}%
\bibitem{Csernai:1999nf}%
  \BibitemOpen
  \bibfield{author}{%
  \bibinfo {author} {\bibfnamefont{L.~P.}\ \bibnamefont{Csernai}}\ and\
  \bibinfo {author} {\bibfnamefont{D.}~\bibnamefont{Rohrich}},\ }%
  \bibfield{journal}{%
  \Doi{10.1016/S0370-2693(99)00615-2}{\bibinfo {journal} {Phys. Lett.}}\ }%
  \textbf{\bibinfo {volume} {B458}},\ \bibinfo {pages} {454} (\bibinfo {year}
  {1999})%
  \bibAnnoteFile{NoStop}{Csernai:1999nf}%
\bibitem{Snellings:1999bt}%
  \BibitemOpen
  \bibfield{author}{%
  \bibinfo {author} {\bibfnamefont{R.~J.~M.}\ \bibnamefont{Snellings}},
  \bibinfo {author} {\bibfnamefont{H.}~\bibnamefont{Sorge}}, \bibinfo {author}
  {\bibfnamefont{S.~A.}\ \bibnamefont{Voloshin}}, \bibinfo {author}
  {\bibfnamefont{F.~Q.}\ \bibnamefont{Wang}},\ and\ \bibinfo {author}
  {\bibfnamefont{N.}~\bibnamefont{Xu}},\ }%
  \bibfield{journal}{%
  \Doi{10.1103/PhysRevLett.84.2803}{\bibinfo {journal} {Phys. Rev. Lett.}}\ }%
  \textbf{\bibinfo {volume} {84}},\ \bibinfo {pages} {2803} (\bibinfo {year}
  {2000})%
  \bibAnnoteFile{NoStop}{Snellings:1999bt}%
\bibitem{Lisa:2000ip}%
  \BibitemOpen
  \bibfield{author}{%
  \bibinfo {author} {\bibfnamefont{M.~A.}\ \bibnamefont{Lisa}}, \bibinfo
  {author} {\bibfnamefont{U.~W.}\ \bibnamefont{Heinz}},\ and\ \bibinfo {author}
  {\bibfnamefont{U.~A.}\ \bibnamefont{Wiedemann}},\ }%
  \bibfield{journal}{%
  \Doi{10.1016/S0370-2693(00)00952-7}{\bibinfo {journal} {Phys. Lett.}}\ }%
  \textbf{\bibinfo {volume} {B489}},\ \bibinfo {pages} {287} (\bibinfo {year}
  {2000})%
  \bibAnnoteFile{NoStop}{Lisa:2000ip}%
\bibitem{Adil:2005qn}%
  \BibitemOpen
  \bibfield{author}{%
  \bibinfo {author} {\bibfnamefont{A.}~\bibnamefont{Adil}}\ and\ \bibinfo
  {author} {\bibfnamefont{M.}~\bibnamefont{Gyulassy}},\ }%
  \bibfield{journal}{%
  \Doi{10.1103/PhysRevC.72.034907}{\bibinfo {journal} {Phys. Rev.}}\ }%
  \textbf{\bibinfo {volume} {C72}},\ \bibinfo {pages} {034907} (\bibinfo {year}
  {2005})%
  \bibAnnoteFile{NoStop}{Adil:2005qn}%
\bibitem{Bozek:2010bi}%
  \BibitemOpen
  \bibfield{author}{%
  \bibinfo {author} {\bibfnamefont{P.}~\bibnamefont{Bo{\.z}ek}}\ and\ \bibinfo
  {author} {\bibfnamefont{I.}~\bibnamefont{Wyskiel}},\ }%
  \bibfield{journal}{%
  \Doi{10.1103/PhysRevC.81.054902}{\bibinfo {journal} {Phys. Rev.}}\ }%
  \textbf{\bibinfo {volume} {C81}},\ \bibinfo {pages} {054902} (\bibinfo {year}
  {2010})%
  \bibAnnoteFile{NoStop}{Bozek:2010bi}%
\bibitem{Steinheimer:2014pfa}%
  \BibitemOpen
  \bibfield{author}{%
  \bibinfo {author} {\bibfnamefont{J.}~\bibnamefont{Steinheimer}}, \bibinfo
  {author} {\bibfnamefont{J.}~\bibnamefont{Auvinen}}, \bibinfo {author}
  {\bibfnamefont{H.}~\bibnamefont{Petersen}}, \bibinfo {author}
  {\bibfnamefont{M.}~\bibnamefont{Bleicher}},\ and\ \bibinfo {author}
  {\bibfnamefont{H.}~\bibnamefont{St{\"o}cker}},\ }%
  \bibfield{journal}{%
  \Doi{10.1103/PhysRevC.89.054913}{\bibinfo {journal} {Phys. Rev.}}\ }%
  \textbf{\bibinfo {volume} {C89}},\ \bibinfo {pages} {054913} (\bibinfo {year}
  {2014})%
  \bibAnnoteFile{NoStop}{Steinheimer:2014pfa}%
\bibitem{Back:2005pc}%
  \BibitemOpen
  \bibfield{author}{%
  \bibinfo {author} {\bibfnamefont{B.~B.}\ \bibnamefont{Back}} \emph{et~al.}
  (\bibinfo {collaboration} {PHOBOS Collaboration}),\ }%
  \bibfield{journal}{%
  \Doi{10.1103/PhysRevLett.97.012301}{\bibinfo {journal} {Phys. Rev. Lett.}}\
  }%
  \textbf{\bibinfo {volume} {97}},\ \bibinfo {pages} {012301} (\bibinfo {year}
  {2006})%
  \bibAnnoteFile{NoStop}{Back:2005pc}%
\bibitem{Abelev:2008jga}%
  \BibitemOpen
  \bibfield{author}{%
  \bibinfo {author} {\bibfnamefont{B.~I.}\ \bibnamefont{Abelev}} \emph{et~al.}
  (\bibinfo {collaboration} {STAR Collaboration}),\ }%
  \bibfield{journal}{%
  \Doi{10.1103/PhysRevLett.101.252301}{\bibinfo {journal} {Phys. Rev. Lett.}}\
  }%
  \textbf{\bibinfo {volume} {101}},\ \bibinfo {pages} {252301} (\bibinfo {year}
  {2008})%
  \bibAnnoteFile{NoStop}{Abelev:2008jga}%
\bibitem{Abelev:2013cva}%
  \BibitemOpen
  \bibfield{author}{%
  \bibinfo {author} {\bibfnamefont{B.}~\bibnamefont{Abelev}} \emph{et~al.}
  (\bibinfo {collaboration} {ALICE}),\ }%
  \bibfield{journal}{%
  \Doi{10.1103/PhysRevLett.111.232302}{\bibinfo {journal} {Phys. Rev. Lett.}}\
  }%
  \textbf{\bibinfo {volume} {111}},\ \bibinfo {pages} {232302} (\bibinfo {year}
  {2013})%
  \bibAnnoteFile{NoStop}{Abelev:2013cva}%
\bibitem{Singha:2016mna}%
  \BibitemOpen
  \bibfield{author}{%
  \bibinfo {author} {\bibfnamefont{S.}~\bibnamefont{Singha}}, \bibinfo {author}
  {\bibfnamefont{P.}~\bibnamefont{Shanmuganathan}},\ and\ \bibinfo {author}
  {\bibfnamefont{D.}~\bibnamefont{Keane}},\ }%
  \bibfield{journal}{%
  \Doi{10.1155/2016/2836989}{\bibinfo {journal} {Adv. High Energy Phys.}}\ }%
  \textbf{\bibinfo {volume} {2016}},\ \bibinfo {pages} {2836989} (\bibinfo
  {year} {2016})%
  \bibAnnoteFile{NoStop}{Singha:2016mna}%
\bibitem{Chatterjee:2017ahy}%
  \BibitemOpen
  \bibfield{author}{%
  \bibinfo {author} {\bibfnamefont{S.}~\bibnamefont{Chatterjee}}\ and\ \bibinfo
  {author} {\bibfnamefont{P.}~\bibnamefont{Bo{\.z}ek}},\ }%
  \bibfield{journal}{%
  \Doi{10.1103/PhysRevLett.120.192301}{\bibinfo {journal} {Phys. Rev. Lett.}}\
  }%
  \textbf{\bibinfo {volume} {120}},\ \bibinfo {pages} {192301} (\bibinfo {year}
  {2018})%
  \bibAnnoteFile{NoStop}{Chatterjee:2017ahy}%
\bibitem{Nasim:2018hyw}%
  \BibitemOpen
  \bibfield{author}{%
  \bibinfo {author} {\bibfnamefont{M.}~\bibnamefont{Nasim}}\ and\ \bibinfo
  {author} {\bibfnamefont{S.}~\bibnamefont{Singha}},\ }%
  \bibfield{journal}{%
  \Doi{10.1103/PhysRevC.97.064917}{\bibinfo {journal} {Phys. Rev.}}\ }%
  \textbf{\bibinfo {volume} {C97}},\ \bibinfo {pages} {064917} (\bibinfo {year}
  {2018})%
  \bibAnnoteFile{NoStop}{Nasim:2018hyw}%
\bibitem{Gursoy:2014aka}%
  \BibitemOpen
  \bibfield{author}{%
  \bibinfo {author} {\bibfnamefont{U.}~\bibnamefont{Gursoy}}, \bibinfo {author}
  {\bibfnamefont{D.}~\bibnamefont{Kharzeev}},\ and\ \bibinfo {author}
  {\bibfnamefont{K.}~\bibnamefont{Rajagopal}},\ }%
  \bibfield{journal}{%
  \Doi{10.1103/PhysRevC.89.054905}{\bibinfo {journal} {Phys. Rev.}}\ }%
  \textbf{\bibinfo {volume} {C89}},\ \bibinfo {pages} {054905} (\bibinfo {year}
  {2014})%
  \bibAnnoteFile{NoStop}{Gursoy:2014aka}%
\bibitem{Das:2016cwd}%
  \BibitemOpen
  \bibfield{author}{%
  \bibinfo {author} {\bibfnamefont{S.~K.}\ \bibnamefont{Das}}, \bibinfo
  {author} {\bibfnamefont{S.}~\bibnamefont{Plumari}}, \bibinfo {author}
  {\bibfnamefont{S.}~\bibnamefont{Chatterjee}}, \bibinfo {author}
  {\bibfnamefont{J.}~\bibnamefont{Alam}}, \bibinfo {author}
  {\bibfnamefont{F.}~\bibnamefont{Scardina}},\ and\ \bibinfo {author}
  {\bibfnamefont{V.}~\bibnamefont{Greco}},\ }%
  \bibfield{journal}{%
  \Doi{10.1016/j.physletb.2017.02.046}{\bibinfo {journal} {Phys. Lett.}}\ }%
  \textbf{\bibinfo {volume} {B768}},\ \bibinfo {pages} {260} (\bibinfo {year}
  {2017})%
  \bibAnnoteFile{NoStop}{Das:2016cwd}%
\bibitem{Brodsky:1977de}%
  \BibitemOpen
  \bibfield{author}{%
  \bibinfo {author} {\bibfnamefont{S.~J.}\ \bibnamefont{Brodsky}}, \bibinfo
  {author} {\bibfnamefont{J.~F.}\ \bibnamefont{Gunion}},\ and\ \bibinfo
  {author} {\bibfnamefont{J.~H.}\ \bibnamefont{Kuhn}},\ }%
  \bibfield{journal}{%
  \Doi{10.1103/PhysRevLett.39.1120}{\bibinfo {journal} {Phys. Rev. Lett.}}\ }%
  \textbf{\bibinfo {volume} {39}},\ \bibinfo {pages} {1120} (\bibinfo {year}
  {1977})%
  \bibAnnoteFile{NoStop}{Brodsky:1977de}%
\bibitem{Bialas:2004su}%
  \BibitemOpen
  \bibfield{author}{%
  \bibinfo {author} {\bibfnamefont{A.}~\bibnamefont{Bialas}}\ and\ \bibinfo
  {author} {\bibfnamefont{W.}~\bibnamefont{Czyz}},\ }%
  \bibfield{journal}{%
  \bibinfo {journal} {Acta Phys. Polon.}\ }%
  \textbf{\bibinfo {volume} {B36}},\ \bibinfo {pages} {905} (\bibinfo {year}
  {2005})%
  \bibAnnoteFile{NoStop}{Bialas:2004su}%
\bibitem{Chesler:2013urd}%
  \BibitemOpen
  \bibfield{author}{%
  \bibinfo {author} {\bibfnamefont{P.~M.}\ \bibnamefont{Chesler}}, \bibinfo
  {author} {\bibfnamefont{M.}~\bibnamefont{Lekaveckas}},\ and\ \bibinfo
  {author} {\bibfnamefont{K.}~\bibnamefont{Rajagopal}},\ }%
  \bibfield{journal}{%
  \Doi{10.1007/JHEP10(2013)013}{\bibinfo {journal} {JHEP}}\ }%
  \textbf{\bibinfo {volume} {10}},\ \bibinfo {pages} {013} (\bibinfo {year}
  {2013})%
  \bibAnnoteFile{NoStop}{Chesler:2013urd}%
\bibitem{Das:2015aga}%
  \BibitemOpen
  \bibfield{author}{%
  \bibinfo {author} {\bibfnamefont{S.~K.}\ \bibnamefont{Das}}, \bibinfo
  {author} {\bibfnamefont{M.}~\bibnamefont{Ruggieri}}, \bibinfo {author}
  {\bibfnamefont{S.}~\bibnamefont{Mazumder}}, \bibinfo {author}
  {\bibfnamefont{V.}~\bibnamefont{Greco}},\ and\ \bibinfo {author}
  {\bibfnamefont{J.-e.}\ \bibnamefont{Alam}},\ }%
  \bibfield{journal}{%
  \Doi{10.1088/0954-3899/42/9/095108}{\bibinfo {journal} {J. Phys.}}\ }%
  \textbf{\bibinfo {volume} {G42}},\ \bibinfo {pages} {095108} (\bibinfo {year}
  {2015})%
  \bibAnnoteFile{NoStop}{Das:2015aga}%
\bibitem{Karpenko:2013wva}%
  \BibitemOpen
  \bibfield{author}{%
  \bibinfo {author} {\bibfnamefont{I.}~\bibnamefont{Karpenko}}, \bibinfo
  {author} {\bibfnamefont{P.}~\bibnamefont{Huovinen}},\ and\ \bibinfo {author}
  {\bibfnamefont{M.}~\bibnamefont{Bleicher}},\ }%
  \bibfield{journal}{%
  \Doi{10.1016/j.cpc.2014.07.010}{\bibinfo {journal} {Comput. Phys. Commun.}}\
  }%
  \textbf{\bibinfo {volume} {185}},\ \bibinfo {pages} {3016} (\bibinfo {year}
  {2014})%
  \bibAnnoteFile{NoStop}{Karpenko:2013wva}%
\bibitem{Chojnacki:2011hb}%
  \BibitemOpen
  \bibfield{author}{%
  \bibinfo {author} {\bibfnamefont{M.}~\bibnamefont{Chojnacki}}, \bibinfo
  {author} {\bibfnamefont{A.}~\bibnamefont{Kisiel}}, \bibinfo {author}
  {\bibfnamefont{W.}~\bibnamefont{Florkowski}},\ and\ \bibinfo {author}
  {\bibfnamefont{W.}~\bibnamefont{Broniowski}},\ }%
  \bibfield{journal}{%
  \Doi{10.1016/j.cpc.2011.11.018}{\bibinfo {journal} {Comput. Phys. Commun.}}\
  }%
  \textbf{\bibinfo {volume} {183}},\ \bibinfo {pages} {746} (\bibinfo {year}
  {2012})%
  \bibAnnoteFile{NoStop}{Chojnacki:2011hb}%
\bibitem{Bozek:2011ua}%
  \BibitemOpen
  \bibfield{author}{%
  \bibinfo {author} {\bibfnamefont{P.}~\bibnamefont{Bo\.zek}},\ }%
  \bibfield{journal}{%
  \Doi{10.1103/PhysRevC.85.034901}{\bibinfo {journal} {Phys. Rev.}}\ }%
  \textbf{\bibinfo {volume} {C85}},\ \bibinfo {pages} {034901} (\bibinfo {year}
  {2012})%
  \bibAnnoteFile{NoStop}{Bozek:2011ua}%
\bibitem{Bozek:2012qs}%
  \BibitemOpen
  \bibfield{author}{%
  \bibinfo {author} {\bibfnamefont{P.}~\bibnamefont{Bo\.zek}}\ and\ \bibinfo
  {author} {\bibfnamefont{I.}~\bibnamefont{Wyskiel-Piekarska}},\ }%
  \bibfield{journal}{%
  \bibinfo {journal} {Phys. Rev.}\ }%
  \textbf{\bibinfo {volume} {C85}},\ \bibinfo {pages} {064915} (\bibinfo {year}
  {2012})%
  \bibAnnoteFile{NoStop}{Bozek:2012qs}%
\bibitem{Sjostrand:2006za}%
  \BibitemOpen
  \bibfield{author}{%
  \bibinfo {author} {\bibfnamefont{T.}~\bibnamefont{Sjostrand}}, \bibinfo
  {author} {\bibfnamefont{S.}~\bibnamefont{Mrenna}},\ and\ \bibinfo {author}
  {\bibfnamefont{P.~Z.}\ \bibnamefont{Skands}},\ }%
  \bibfield{journal}{%
  \Doi{10.1088/1126-6708/2006/05/026}{\bibinfo {journal} {JHEP}}\ }%
  \textbf{\bibinfo {volume} {05}},\ \bibinfo {pages} {026} (\bibinfo {year}
  {2006})%
  \bibAnnoteFile{NoStop}{Sjostrand:2006za}%
\bibitem{Sjostrand:2007gs}%
  \BibitemOpen
  \bibfield{author}{%
  \bibinfo {author} {\bibfnamefont{T.}~\bibnamefont{Sjostrand}}, \bibinfo
  {author} {\bibfnamefont{S.}~\bibnamefont{Mrenna}},\ and\ \bibinfo {author}
  {\bibfnamefont{P.~Z.}\ \bibnamefont{Skands}},\ }%
  \bibfield{journal}{%
  \Doi{10.1016/j.cpc.2008.01.036}{\bibinfo {journal} {Comput. Phys. Commun.}}\
  }%
  \textbf{\bibinfo {volume} {178}},\ \bibinfo {pages} {852} (\bibinfo {year}
  {2008})%
  \bibAnnoteFile{NoStop}{Sjostrand:2007gs}%
\bibitem{He:2013zua}%
  \BibitemOpen
  \bibfield{author}{%
  \bibinfo {author} {\bibfnamefont{M.}~\bibnamefont{He}}, \bibinfo {author}
  {\bibfnamefont{H.}~\bibnamefont{van Hees}}, \bibinfo {author}
  {\bibfnamefont{P.~B.}\ \bibnamefont{Gossiaux}}, \bibinfo {author}
  {\bibfnamefont{R.~J.}\ \bibnamefont{Fries}},\ and\ \bibinfo {author}
  {\bibfnamefont{R.}~\bibnamefont{Rapp}},\ }%
  \bibfield{journal}{%
  \Doi{10.1103/PhysRevE.88.032138}{\bibinfo {journal} {Phys. Rev.}}\ }%
  \textbf{\bibinfo {volume} {E88}},\ \bibinfo {pages} {032138} (\bibinfo {year}
  {2013})%
  \bibAnnoteFile{NoStop}{He:2013zua}%
\bibitem{Peterson:1982ak}%
  \BibitemOpen
  \bibfield{author}{%
  \bibinfo {author} {\bibfnamefont{C.}~\bibnamefont{Peterson}}, \bibinfo
  {author} {\bibfnamefont{D.}~\bibnamefont{Schlatter}}, \bibinfo {author}
  {\bibfnamefont{I.}~\bibnamefont{Schmitt}},\ and\ \bibinfo {author}
  {\bibfnamefont{P.~M.}\ \bibnamefont{Zerwas}},\ }%
  \bibfield{journal}{%
  \Doi{10.1103/PhysRevD.27.105}{\bibinfo {journal} {Phys. Rev.}}\ }%
  \textbf{\bibinfo {volume} {D27}},\ \bibinfo {pages} {105} (\bibinfo {year}
  {1983})%
  \bibAnnoteFile{NoStop}{Peterson:1982ak}%
\bibitem{Tuchin:2013apa}%
  \BibitemOpen
  \bibfield{author}{%
  \bibinfo {author} {\bibfnamefont{K.}~\bibnamefont{Tuchin}},\ }%
  \bibfield{journal}{%
  \Doi{10.1103/PhysRevC.88.024911}{\bibinfo {journal} {Phys. Rev.}}\ }%
  \textbf{\bibinfo {volume} {C88}},\ \bibinfo {pages} {024911} (\bibinfo {year}
  {2013})%
  \bibAnnoteFile{NoStop}{Tuchin:2013apa}%
\bibitem{vanHees:2004gq}%
  \BibitemOpen
  \bibfield{author}{%
  \bibinfo {author} {\bibfnamefont{H.}~\bibnamefont{van Hees}}\ and\ \bibinfo
  {author} {\bibfnamefont{R.}~\bibnamefont{Rapp}},\ }%
  \bibfield{journal}{%
  \Doi{10.1103/PhysRevC.71.034907}{\bibinfo {journal} {Phys. Rev.}}\ }%
  \textbf{\bibinfo {volume} {C71}},\ \bibinfo {pages} {034907} (\bibinfo {year}
  {2005})%
  \bibAnnoteFile{NoStop}{vanHees:2004gq}%
\bibitem{Moore:2004tg}%
  \BibitemOpen
  \bibfield{author}{%
  \bibinfo {author} {\bibfnamefont{G.~D.}\ \bibnamefont{Moore}}\ and\ \bibinfo
  {author} {\bibfnamefont{D.}~\bibnamefont{Teaney}},\ }%
  \bibfield{journal}{%
  \Doi{10.1103/PhysRevC.71.064904}{\bibinfo {journal} {Phys. Rev.}}\ }%
  \textbf{\bibinfo {volume} {C71}},\ \bibinfo {pages} {064904} (\bibinfo {year}
  {2005})%
  \bibAnnoteFile{NoStop}{Moore:2004tg}%
\bibitem{Gubser:2006qh}%
  \BibitemOpen
  \bibfield{author}{%
  \bibinfo {author} {\bibfnamefont{S.~S.}\ \bibnamefont{Gubser}},\ }%
  \bibfield{journal}{%
  \Doi{10.1103/PhysRevD.76.126003}{\bibinfo {journal} {Phys. Rev.}}\ }%
  \textbf{\bibinfo {volume} {D76}},\ \bibinfo {pages} {126003} (\bibinfo {year}
  {2007})%
  \bibAnnoteFile{NoStop}{Gubser:2006qh}%
\bibitem{Alberico:2011zy}%
  \BibitemOpen
  \bibfield{author}{%
  \bibinfo {author} {\bibfnamefont{W.~M.}\ \bibnamefont{Alberico}}, \bibinfo
  {author} {\bibfnamefont{A.}~\bibnamefont{Beraudo}}, \bibinfo {author}
  {\bibfnamefont{A.}~\bibnamefont{De~Pace}}, \bibinfo {author}
  {\bibfnamefont{A.}~\bibnamefont{Molinari}}, \bibinfo {author}
  {\bibfnamefont{M.}~\bibnamefont{Monteno}}, \bibinfo {author}
  {\bibfnamefont{M.}~\bibnamefont{Nardi}},\ and\ \bibinfo {author}
  {\bibfnamefont{F.}~\bibnamefont{Prino}},\ }%
  \bibfield{journal}{%
  \Doi{10.1140/epjc/s10052-011-1666-6}{\bibinfo {journal} {Eur. Phys. J.}}\ }%
  \textbf{\bibinfo {volume} {C71}},\ \bibinfo {pages} {1666} (\bibinfo {year}
  {2011})%
  \bibAnnoteFile{NoStop}{Alberico:2011zy}%
\bibitem{Berrehrah:2014tva}%
  \BibitemOpen
  \bibfield{author}{%
  \bibinfo {author} {\bibfnamefont{H.}~\bibnamefont{Berrehrah}}, \bibinfo
  {author} {\bibfnamefont{P.~B.}\ \bibnamefont{Gossiaux}}, \bibinfo {author}
  {\bibfnamefont{J.}~\bibnamefont{Aichelin}}, \bibinfo {author}
  {\bibfnamefont{W.}~\bibnamefont{Cassing}}, \bibinfo {author}
  {\bibfnamefont{J.~M.}\ \bibnamefont{Torres-Rincon}},\ and\ \bibinfo {author}
  {\bibfnamefont{E.}~\bibnamefont{Bratkovskaya}},\ }%
  \bibfield{journal}{%
  \Doi{10.1103/PhysRevC.90.051901}{\bibinfo {journal} {Phys. Rev.}}\ }%
  \textbf{\bibinfo {volume} {C90}},\ \bibinfo {pages} {051901} (\bibinfo {year}
  {2014})%
  \bibAnnoteFile{NoStop}{Berrehrah:2014tva}%
\bibitem{Scardina:2017ipo}%
  \BibitemOpen
  \bibfield{author}{%
  \bibinfo {author} {\bibfnamefont{F.}~\bibnamefont{Scardina}}, \bibinfo
  {author} {\bibfnamefont{S.~K.}\ \bibnamefont{Das}}, \bibinfo {author}
  {\bibfnamefont{V.}~\bibnamefont{Minissale}}, \bibinfo {author}
  {\bibfnamefont{S.}~\bibnamefont{Plumari}},\ and\ \bibinfo {author}
  {\bibfnamefont{V.}~\bibnamefont{Greco}},\ }%
  \bibfield{journal}{%
  \Doi{10.1103/PhysRevC.96.044905}{\bibinfo {journal} {Phys. Rev.}}\ }%
  \textbf{\bibinfo {volume} {C96}},\ \bibinfo {pages} {044905} (\bibinfo {year}
  {2017})%
  \bibAnnoteFile{NoStop}{Scardina:2017ipo}%
\bibitem{Xu:2017obm}%
  \BibitemOpen
  \bibfield{author}{%
  \bibinfo {author} {\bibfnamefont{Y.}~\bibnamefont{Xu}}, \bibinfo {author}
  {\bibfnamefont{J.~E.}\ \bibnamefont{Bernhard}}, \bibinfo {author}
  {\bibfnamefont{S.~A.}\ \bibnamefont{Bass}}, \bibinfo {author}
  {\bibfnamefont{M.}~\bibnamefont{Nahrgang}},\ and\ \bibinfo {author}
  {\bibfnamefont{S.}~\bibnamefont{Cao}},\ }%
  \bibfield{journal}{%
  \Doi{10.1103/PhysRevC.97.014907}{\bibinfo {journal} {Phys. Rev.}}\ }%
  \textbf{\bibinfo {volume} {C97}},\ \bibinfo {pages} {014907} (\bibinfo {year}
  {2018})%
  \bibAnnoteFile{NoStop}{Xu:2017obm}%
\bibitem{Greco:2003vf}%
  \BibitemOpen
  \bibfield{author}{%
  \bibinfo {author} {\bibfnamefont{V.}~\bibnamefont{Greco}}, \bibinfo {author}
  {\bibfnamefont{C.~M.}\ \bibnamefont{Ko}},\ and\ \bibinfo {author}
  {\bibfnamefont{R.}~\bibnamefont{Rapp}},\ }%
  \bibfield{journal}{%
  \Doi{10.1016/j.physletb.2004.06.064}{\bibinfo {journal} {Phys. Lett.}}\ }%
  \textbf{\bibinfo {volume} {B595}},\ \bibinfo {pages} {202} (\bibinfo {year}
  {2004})%
  \bibAnnoteFile{NoStop}{Greco:2003vf}%
\bibitem{vanHees:2005wb}%
  \BibitemOpen
  \bibfield{author}{%
  \bibinfo {author} {\bibfnamefont{H.}~\bibnamefont{van Hees}}, \bibinfo
  {author} {\bibfnamefont{V.}~\bibnamefont{Greco}},\ and\ \bibinfo {author}
  {\bibfnamefont{R.}~\bibnamefont{Rapp}},\ }%
  \bibfield{journal}{%
  \Doi{10.1103/PhysRevC.73.034913}{\bibinfo {journal} {Phys. Rev.}}\ }%
  \textbf{\bibinfo {volume} {C73}},\ \bibinfo {pages} {034913} (\bibinfo {year}
  {2006})%
  \bibAnnoteFile{NoStop}{vanHees:2005wb}%
\bibitem{Cao:2013ita}%
  \BibitemOpen
  \bibfield{author}{%
  \bibinfo {author} {\bibfnamefont{S.}~\bibnamefont{Cao}}, \bibinfo {author}
  {\bibfnamefont{G.-Y.}\ \bibnamefont{Qin}},\ and\ \bibinfo {author}
  {\bibfnamefont{S.~A.}\ \bibnamefont{Bass}},\ }%
  \bibfield{journal}{%
  \Doi{10.1103/PhysRevC.88.044907}{\bibinfo {journal} {Phys. Rev.}}\ }%
  \textbf{\bibinfo {volume} {C88}},\ \bibinfo {pages} {044907} (\bibinfo {year}
  {2013})%
  \bibAnnoteFile{NoStop}{Cao:2013ita}%
\bibitem{Song:2015ykw}%
  \BibitemOpen
  \bibfield{author}{%
  \bibinfo {author} {\bibfnamefont{T.}~\bibnamefont{Song}}, \bibinfo {author}
  {\bibfnamefont{H.}~\bibnamefont{Berrehrah}}, \bibinfo {author}
  {\bibfnamefont{D.}~\bibnamefont{Cabrera}}, \bibinfo {author}
  {\bibfnamefont{W.}~\bibnamefont{Cassing}},\ and\ \bibinfo {author}
  {\bibfnamefont{E.}~\bibnamefont{Bratkovskaya}},\ }%
  \bibfield{journal}{%
  \Doi{10.1103/PhysRevC.93.034906}{\bibinfo {journal} {Phys. Rev.}}\ }%
  \textbf{\bibinfo {volume} {C93}},\ \bibinfo {pages} {034906} (\bibinfo {year}
  {2016})%
  \bibAnnoteFile{NoStop}{Song:2015ykw}%
\bibitem{Plumari:2017ntm}%
  \BibitemOpen
  \bibfield{author}{%
  \bibinfo {author} {\bibfnamefont{S.}~\bibnamefont{Plumari}}, \bibinfo
  {author} {\bibfnamefont{V.}~\bibnamefont{Minissale}}, \bibinfo {author}
  {\bibfnamefont{S.~K.}\ \bibnamefont{Das}}, \bibinfo {author}
  {\bibfnamefont{G.}~\bibnamefont{Coci}},\ and\ \bibinfo {author}
  {\bibfnamefont{V.}~\bibnamefont{Greco}},\ }%
  \bibfield{journal}{%
  \Doi{10.1140/epjc/s10052-018-5828-7}{\bibinfo {journal} {Eur. Phys. J.}}\ }%
  \textbf{\bibinfo {volume} {C78}},\ \bibinfo {pages} {348} (\bibinfo {year}
  {2018})%
  \bibAnnoteFile{NoStop}{Plumari:2017ntm}%
\bibitem{Das:2015ana}%
  \BibitemOpen
  \bibfield{author}{%
  \bibinfo {author} {\bibfnamefont{S.~K.}\ \bibnamefont{Das}}, \bibinfo
  {author} {\bibfnamefont{F.}~\bibnamefont{Scardina}}, \bibinfo {author}
  {\bibfnamefont{S.}~\bibnamefont{Plumari}},\ and\ \bibinfo {author}
  {\bibfnamefont{V.}~\bibnamefont{Greco}},\ }%
  \bibfield{journal}{%
  \Doi{10.1016/j.physletb.2015.06.003}{\bibinfo {journal} {Phys. Lett.}}\ }%
  \textbf{\bibinfo {volume} {B747}},\ \bibinfo {pages} {260} (\bibinfo {year}
  {2015})%
  \bibAnnoteFile{NoStop}{Das:2015ana}%
\bibitem{Cao:2015hia}%
  \BibitemOpen
  \bibfield{author}{%
  \bibinfo {author} {\bibfnamefont{S.}~\bibnamefont{Cao}}, \bibinfo {author}
  {\bibfnamefont{G.-Y.}\ \bibnamefont{Qin}},\ and\ \bibinfo {author}
  {\bibfnamefont{S.~A.}\ \bibnamefont{Bass}},\ }%
  \bibfield{journal}{%
  \Doi{10.1103/PhysRevC.92.024907}{\bibinfo {journal} {Phys. Rev.}}\ }%
  \textbf{\bibinfo {volume} {C92}},\ \bibinfo {pages} {024907} (\bibinfo {year}
  {2015})%
  \bibAnnoteFile{NoStop}{Cao:2015hia}%
\bibitem{Singha:2018cdj}%
  \BibitemOpen
  \bibfield{author}{%
  \bibinfo {author} {\bibfnamefont{S.}~\bibnamefont{Singha}} (\bibinfo
  {collaboration} {STAR Collaboration}),\ }%
  \bibfield{journal}{%
  \Doi{10.1016/j.nuclphysa.2018.09.010}{\bibinfo {journal} {Nucl. Phys.}}\ }%
  \textbf{\bibinfo {volume} {A982}},\ \bibinfo {pages} {671} (\bibinfo {year}
  {2019})%
  \bibAnnoteFile{NoStop}{Singha:2018cdj}%
\bibitem{Grosa:2018zix}%
  \BibitemOpen
  \bibfield{author}{%
  \bibinfo {author} {\bibfnamefont{F.}~\bibnamefont{Grosa}} (\bibinfo
  {collaboration} {ALICE Collaboration}),\ }%
  \bibfield{journal}{%
  \Doi{10.22323/1.345.0138}{\bibinfo {journal} {PoS}}\ }%
  \textbf{\bibinfo {volume} {HardProbes2018}},\ \bibinfo {pages} {138}
  (\bibinfo {year} {2018})%
  \bibAnnoteFile{NoStop}{Grosa:2018zix}%
\end{thebibliography}%

\end{document}